\documentclass{aa}
\usepackage[varg]{txfonts}

\usepackage{natbib}
\bibpunct{(}{)}{;}{a}{}{,}
\usepackage[colorlinks=true,citecolor=blue]{hyperref}

\usepackage{longtable}
\usepackage{lscape}

\usepackage{verbatim}
\usepackage{makecell}
\usepackage{multirow}
\usepackage{booktabs}

\begin{document}

\title{Analysis of high-resolution FEROS spectroscopy for a sample of variable B-type stars assembled from TESS photometry\thanks{Based on observations collected at the European Southern Observatory, La Silla, Chile under program 0104.A-9001(A).}\fnmsep\thanks{Tables A.1 and A.2 are only available in electronic form at the CDS via anonymous ftp to cdsarc.u-strasbg.fr (130.79.128.5) or via \url{http://cdsweb.u-strasbg.fr/cgi-bin/qcat?J/A+A/}}}

\author{Sarah Gebruers\inst{\ref{KUL},\ref{MPIA}}
\and Andrew Tkachenko\inst{\ref{KUL}}
\and Dominic M. Bowman\inst{\ref{KUL}} 
\and Timothy Van Reeth\inst{\ref{KUL}} 
\and Siemen Burssens\inst{\ref{KUL}} 
\and Luc IJspeert\inst{\ref{KUL}} 
\and Laurent Mahy\inst{\ref{ROB}}
\and Ilya Straumit\inst{\ref{KUL},\ref{Ohio}}
\and Maosheng Xiang\inst{\ref{MPIA}}
\and Hans-Walter Rix\inst{\ref{MPIA}}
\and Conny Aerts\inst{\ref{KUL},\ref{MPIA},\ref{Radboud}}
}  
\institute{Institute of Astronomy, KU Leuven, Celestijnenlaan 200D, B-3001 Leuven, Belgium \\ \email{sarah.gebruers@kuleuven.be} \label{KUL} 
\and Max Planck Institute for Astronomy, K\"onigstuhl 17, 69117 Heidelberg, Germany \label{MPIA}
\and Koninklijke Sterrenwacht van België, Ringlaan 3, B-1180 Brussel, België \label{ROB} 
\and The Department of Astronomy and Center of Cosmology and AstroParticle Physics, The Ohio State University, Columbus, OH 43210, USA \label{Ohio}
\and Department of Astrophysics, IMAPP, Radboud University Nijmegen, PO Box 9010, 6500 GL Nijmegen, The Netherlands\label{Radboud}
}

\date{Received XXX / Accepted XXX}

\abstract{Spectroscopic data are necessary to break degeneracies in asteroseismic modelling of the interior structure in high- and intermediate-mass stars. With the TESS mission, the number of bright intermediate-mass B-type stars with long photometric light curves, that are therefore suitable for detailed asteroseismic studies, has increased substantially compared to the pre-TESS era.} 
{We derive precise photospheric stellar parameters for a sample of 166 B-type stars with TESS light curves through a homogeneous spectroscopic analysis. The variability types of these sample stars are also classified based on all currently available TESS sectors and ultimately prioritised according to their astrophysical potential.}
{We obtained high-resolution spectra for all 166 targets with the FEROS spectrograph in the context of a large program. The spectra are reduced with the CERES pipeline, that we adapted to improve the quality of the reduced spectra. These spectra are subsequently analysed with the {\sc zeta-Payne}, a machine learning-based spectrum analysis algorithm, to infer precise stellar labels for all stars in the sample. Furthermore, the Least-Squares Deconvolution (LSD) method is employed to investigate spectral line profile variability (LPV) and isolate binary systems from presumably single stars.} 
{The LSD profile analysis identified 26 spectroscopic double-lined binaries; the remainder of the sample are 42 supergiants in the Large Magellanic Cloud galaxy and 98 Galactic stars, both with and without apparent LPV. For the Galactic single stars and single-lined spectroscopic binaries, we determine their five main surface parameters: effective temperature ($T_{\mathrm{eff}}$), surface gravity ($\log\,g$), global metallicity ([M/H]), projected rotational velocity ($v\sin\,i$), and microturbulent velocity ($\xi$) with average formal precisions of 70\,K, 0.03\,dex, 0.07\,dex, 8\,km\,s$^{-1}$, and 0.7\,km\,s$^{-1}$ respectively. The average internal uncertainties we find for FEROS spectra with our spectrum analysis method are 430\,K ($T_{\mathrm{eff}}$), 0.12\,dex ($\log\,g$), 0.13\,dex ([M/H]), 12\,km\,s$^{-1}$ ($v\sin\,i$), and 2\,km\,s$^{-1}$ ($\xi$).}
{We find spectroscopic evidence that eight of the 98 galactic single or SB1 variables are fast rotating gravity-mode pulsators occurring in between the slowly pulsating B (SPB) stars and $\delta\,$Scuti instability strips. The g-mode frequencies of these pulsators are shifted to relatively high frequency values due to their rotation, and their apparently too low $T_{\mathrm{eff}}$ relative to the SPB instability region can in most cases be explained by the gravity darkening effect. We also discover 13 new HgMn stars in the Galactic sample of which only one is found in a spectroscopic binary, resulting in a biased and therefore unreliable low binary rate of only 8\%.}


\titlerunning{Analysis of FEROS spectroscopy for a sample of variable B-type stars with TESS photometry}
\authorrunning{Gebruers et al.}
\maketitle

\section{Introduction}

While massive stars are important suppliers of chemical and mechanical feedback to the interstellar medium \citep[e.g.][]{Langer2012}, their structure and evolution is not yet well understood. Large theoretical uncertainties occur already during the early phases, that have an impact on the further evolution \citep[e.g.][]{Maeder&Meynet2000,Ekstrom2012,Aerts2019}. To calibrate physical processes in these stars, observational constraints from their interiors are needed. With space asteroseismology, observed pulsations inside stars provide information about stellar interiors \citep{Aerts2021}. 
Low-mass stars pulsate with pressure modes that probe the stellar envelope. They have a similar structure to that of the Sun, thus similar analysis techniques can be used as in helioseismology \citep{Christensen-Dalsgaard2002}. 
On the other hand, intermediate to massive main-sequence stars (M $\sim$ 2-25\,M$_{\odot}$) pulsate both in pressure (p) and gravity (g) modes, either simultaneously (the case of hybrid pulsators) or separately. Gravity-mode oscillations are most sensitive to physical conditions in the near-core regions of stars with M $>$ 1.3\,M$_\odot$, and are therefore ideal to unravel the mechanisms of convective core boundary mixing, near-core rotation and chemical mixing (see \citealt{Bowman2020c} for a review). 
Many theoretical and observational studies of single stars or their ensembles have used g-mode pulsations to infer internal information such as the profile of the convective boundary mixing region and the radiative envelope, the internal rotation rate, and the amount of chemical mixing, both in slowly pulsating B (SPB) stars and $\beta$\,Cephei ($\beta$\,Cep) stars \citep[e.g.][]{Miglio2008,Szewczuk2015,Moravveji2016,Pedersen2018,Michielsen2019,Mozdzierski2019}, and in less massive $\gamma$ Doradus ($\gamma$ Dor) stars \citep[e.g.][]{VanReeth2018,Mombarg2020,Angelou2022}.
These are three classes of pulsating main-sequence stars with predominantly g-mode pulsations in $\gamma\,$Dor (1.3-1.9\,M$_\odot$) and SPB (3-8\,M$_\odot$) stars and p-mode pulsations in $\beta\,$Cep stars (8-25\,M$_\odot$).

There are not that many asteroseismic studies of massive stars, as compared to low-mass stars, because long time base, continuous and high-precision time series are needed to properly resolve frequencies and identify geometries of g-mode pulsations \citep{Bowman2020c}. This became possible with the start of science operations of space missions such as CoRoT \citep{Auvergne2009} and \textit{Kepler} \citep{Borucki2010}. Especially the \textit{Kepler} mission revolutionised the field of g-mode asteroseismology with its four-year duration light curves. However, \textit{Kepler} was pointed to one fixed field of view for its entire nominal mission duration, where the field selection was driven by the primary science of the mission of detection and characterisation of exoplanets around solar-like stars. Therefore, not many massive stars were observed by the mission.
The ongoing Transiting Exoplanet Survey Satellite (TESS) mission \citep{Ricker2015} is observing the whole sky searching for planets around bright stars, and thereby also targeting O- and B-type stars. The nominal mission divided the two ecliptic hemispheres into 13 sectors that are each observed for 27 d. The 13 sectors in the Northern and Southern ecliptic hemispheres overlap near the poles where one-year long observations are taken. In the Southern hemisphere, the TESS continuous viewing zone (CVZ) includes the Large Magellanic Cloud (LMC) galaxy which means that different metallicity regimes can be probed.
Different studies have demonstrated the diverse variability among OB-type stars using \textit{Kepler}/K2 and TESS light curves. \citet{Bowman2019b} and \cite{Burssens2019} found a number of coherent pulsators among more than 100 OB-type stars in the K2 mission, and light curves dominated by stochastic low-frequency variability (SLF) for the most massive stars and dozens of blue supergiants observed with K2 and TESS. From the first two sectors of TESS, \citet{Pedersen2019} detected and classified the variability for a sample of 154 OB-type stars, including 40 LMC targets, showing that 90\% have pulsational, binary, rotational or SLF variability. \citet{Burssens2020} also found a large diversity of variability in a sample of OB-type stars observed within the first 13 sectors of TESS, and demonstrated the power of coupling TESS light curves with high-resolution spectroscopy.

Asteroseismic modelling of intermediate-mass stars entails a number of degeneracies due to strong correlations between the numerous parameters occurring in stellar models. The addition of spectroscopic, binarity or astrometric data helps to lift these degeneracies, reduce uncertainties on the parameters, and supplies information that is required to establish a link between interiors of stars and their atmospheric layers. With this in mind, many studies in the past decade have focused on obtaining spectroscopic parameters for samples of promising pulsating \textit{Kepler} stars \citep[e.g.][]{Lehmann2011,Tkachenko2012,Tkachenko2013a,Tkachenko2013b,VanReeth2015,Niemczura2015,Niemczura2017}. 
During early phases of the \textit{Kepler} mission science operations, ground-based spectroscopy was frequently used in tandem with \textit{Kepler} photometry to aid with the photometric classification of variable stars. Effective temperature ($T_{\mathrm{eff}}$) and surface gravity ($\log\,g$) estimates were used to place stars in the Kiel diagram ($\log\,g$ as a function of $T_{\mathrm{eff}}$) and compare their positions with theoretical instability strips of different types of pulsators, thus confirming or rejecting the photometric classification \citep[e.g.][]{Uytterhoeven2011,Balona2011,Tkachenko2013b}. The spectroscopic surface measurements were also used to check estimates of those parameters determined from photometry \citep[e.g.][]{SilvaAguirre2012,Thygesen2012,Pinsonneault2014}. The other way around, asteroseismic values of $\log\,g$ are often more precise than spectroscopic ones and have thus been used in some studies to better constrain spectroscopic parameters of stars \citep{Bruntt2012,Thygesen2012}. 

In recent years, detailed asteroseismic modelling of intermediate-mass stars has made advances by including non-photometric data. For example, \citet{Mombarg2019} added spectroscopic measurements of $T_{\mathrm{eff}}$ and $\log\,g$ to the forward modelling of $\gamma$ Dor stars to lift degeneracies and to estimate masses and ages of stars. In \citet{Mombarg2020}, atomic diffusion in AF-type stars was studied by not only adding spectroscopic $T_{\mathrm{eff}}$ and $\log\,g$ values to g-mode asteroseismic modelling, but also by comparing the predicted surface abundances to observational ones to evaluate models with and without atomic diffusion. \citet{Pedersen2018} studied different mixing profiles in more massive B-type stars and emphasised the importance of adding surface abundances to asteroseismic modelling to constrain the shape of the mixing profiles in radiative zones of these stars. \citet{Pedersen2021} performed forward asteroseismic modelling on a sample of 26 SPB stars where they applied spectroscopic parameters and an astrometric measurement of the luminosity to delimit the parameter space. 
Aside from spectroscopy, binary modelling can also add valuable constraints to asteroseismic modelling since it can deliver independent estimates of the radius and mass with very high precision. \citet{Johnston2019}, for example, used binary information to reduce parameter uncertainties and correlations between parameters in the modelling of binaries with g-mode pulsators. \citet{Sekaran2020,Sekaran2021} combined \textit{Kepler} photometry with high-resolution optical {\sc hermes} spectroscopy \citep{Raskin2011} to perform a detailed asteroseismic study of a $\gamma\,$Dor-type g-mode pulsator in an eclipsing double-lined spectroscopic binary (SB2) system. The authors demonstrated that it is highly beneficial for asteroseismic modelling to add model-independent information on stellar mass and radius inferred from binary dynamics. 

Various classes of chemically peculiar (CP) stars have been encountered among intermediate mass B-type stars.
One of them are CP2 or ApBp stars that show overabundances of Sr, Cr and Eu. For a subset of ApBp stars these overabundances, located in spots on the stellar surface, can be explained by the presence of a magnetic field \citep{Smith1996,Gray2005}. ApBp stars are typically slow rotators and they are not likely to be in (close) binary systems \citep{Abt1973,Mathys2020}. 
The other group are CP3 or HgMn stars which have enhanced abundances of Hg and Mn \citep[and other elements such as Ga, Sr and Y,][]{Smith1996}. HgMn stars are slow rotators with $\big\langle v\sin\,i \big\rangle \sim$ 29\,km\,s$^{-1}$ \citep{Abt1972} and have a high binarity rate with spectroscopic binarity percentages between 50-91\% \citep[e.g.][]{Hubrig1995, Scholler2010}. About half of the binaries are found in SB2 systems, mostly with short orbital periods. Although they are thought to be non-magnetic CP stars, they show rotational modulation consistent with spots of overabundant chemical elements on the surface \citep[e.g.][]{Korhonen2013,Kochukhov2021}. Unlike for ApBp stars, the origin of these spots in HgMn stars is not yet fully understood. Possible explanations are atomic diffusion processes in the outer stellar layers \citep{Alecian2011} or maybe weak magnetic fields \citep{Hubrig2020}. Another cause for the variability seen in HgMn stars, aside from rotational modulation, are pulsations but not many pulsating HgMn stars have been detected yet \citep[e.g.][]{Alecian2009,Kochukhov2021}.

In this paper, we analyse the spectra of 166 B-type stars that exhibit photometric variability in their TESS light curves, with the purpose of future combined asteroseismic and spectroscopic modelling. For the sake of efficiency and consistency of the obtained results across the entire stellar sample, we employ the {\sc zeta-Payne} \citep{Straumit2022} machine learning-based spectrum analysis algorithm. The method is a generalisation of the originally proposed {\sc The Payne} algorithm \citep{Ting2019} towards the inclusion of a model with an a priori unknown residual instrumental response function into the parameter vector. This way, parameters of the residual response function are optimised along with the atmospheric parameters of the star, thus removing the need to normalise stellar spectra manually in advance.

In Sect.~\ref{sect:observations}, we discuss the sample selection and observations of the targets. We describe some improvements made to the spectrum reduction pipeline in Sect.~\ref{sect:CERES}. The spectroscopic classification is presented in Sect.~\ref{sect:methods} along with a summary of the {\sc zeta-Payne} framework and a few tests for the specific set-up in this paper. The spectroscopic and photometric results are given in Sect.~\ref{sect:results}, while the internal uncertainties of the methodology and a subsample of peculiar stars are discussed in Sect.~\ref{sect:discussion}. This section also presents the obtained results in the context of the spectroscopic Hertzsprung-Russell diagram \citep{Langer2014}. We conclude our study in Sect.~\ref{sect:conclusion}.

\section{Sample selection and observations} \label{sect:observations}

We selected all the bright targets (stars with V-band magnitudes below 11.5\,mag) in the samples of variable OB-type stars from \citet{Pedersen2019} and \citet{Bowman2019b}. These samples were constructed following the asteroseismic potential of OB-type stars as determined from the first three TESS sectors that were available at that time. 
For the 166 brightest stars, we obtained spectroscopic observations with the Fibre-fed Extended Range Optical Spectrograph \citep[FEROS,][]{Kaufer1999} which is attached to the ESO/MPG 2.2-m telescope at La Silla, Chile. In an accompanying paper (Serebriakova et al., in prep.), the spectroscopic analysis of ESO UVES spectra for fainter stars ($V > 10$\,mag) from the samples of \citet{Pedersen2019} and \citet{Bowman2019b} will be presented. 

FEROS is a high-resolution spectrograph ($R \sim$ 48\,000) that covers a wavelength range from 3600-9200\,\AA \ over 39 orders. The observations were taken during December 2019 and February 2020. In total 166 objects were observed with a minimum of two epochs per target (separated by two months), except for seven targets that were not observable during the second run in February. 
The spectra were reduced with the publicly available Collection of Elemental Routines for Echelle Spectra \citep[CERES,][]{Brahm2017}\footnote{\url{https://github.com/rabrahm/ceres}}. This package consists of automated reduction pipelines for 13 different \'echelle spectrographs, among which FEROS. The pipelines are written in {\tt Python} and complemented with {\tt C} and {\tt Fortran} routines to speed-optimise the time consuming parts. It performs a bias subtraction, correction for bad columns, background subtraction, order extraction, wavelength calibration, barycentric correction and a blaze correction. 
To achieve the goals of this paper, some changes to the CERES pipeline were necessary to obtain most optimally extracted 1D spectra. These changes are described in Sect.~\ref{sect:CERES}. 

To complement the spectroscopic data within this work and to classify the dominant variability type of each target, we re-analysed available 2-min cadence photometry from TESS. We downloaded the available simple aperture photometry (SAP) and pre-data search conditioning (PDC-SAP) light curves output from the NASA SPOC pipeline \citep{Jenkins2016b} from the Mikulsi Archive for Space Telescopes (MAST\footnote{\url{https://archive.stsci.edu/}}). Almost all bright ($V \leq 14$\,mag) O and early-B stars observed by TESS were prioritised for 2-min cadence in at least one sector because of successful guest investigator proposals\footnote{GO3059 and GO4074; PI: Bowman; \url{https://heasarc.gsfc.nasa.gov/docs/tess/ approved- programs.html}}. Yet, some stars fall between the gaps in the CCDs and cameras, or are located close to the ecliptic. Cycle 4 of the TESS mission is currently in progress as of July 2021, which includes the second visit to the northern ecliptic hemisphere and parts of the ecliptic for the first time.

(Almost) all of our targets currently have at least one sector of TESS data available, which is sufficient to classify its dominant variability. Given that variability classification is sufficient for this paper, as its focus is on the analysis of the newly-obtained spectroscopic data for a large number of B-type stars, we focused our photometric analysis on the longer time spans of 2-min TESS light curves available for each target.
We checked both the available SAP and PDC-SAP light curves to compare and contrast the impact of instrumental systematics that may be present in either or both light curve formats. In most cases, the latter were sufficient for variability classification. However, we note that correcting for instrumental effects (for example extracting custom light curves) to derive reliable lists of pulsation frequencies is the subject of future work.

\section{Improvements of the CERES pipeline for the FEROS spectra}\label{sect:CERES}

The CERES pipeline reduces 25 of the 39 FEROS orders, which corresponds to the wavelength range 3800-6800\,\AA, containing six Balmer lines and sufficient metal lines to determine stellar parameters \citep{Brahm2017}.
We made four changes to the existing CERES pipeline in order to meet our quest for high-precision spectroscopic parameter determination of our targets. The first two were necessary to remove unphysical artefacts (hereafter referred to as wiggles) from the spectra and to eliminate `bumps' occurring at the edges of the orders. The other two changes concern an additional function that merges the individual orders and a method to remove cosmic hits. The original CERES pipeline only returns separate orders and not the fully merged spectrum since the main purpose of the developers was to determine precise radial velocities (RV) which is possible from individual orders. 

\begin{figure*}[ht]
    \centering
    \includegraphics[width=\textwidth]{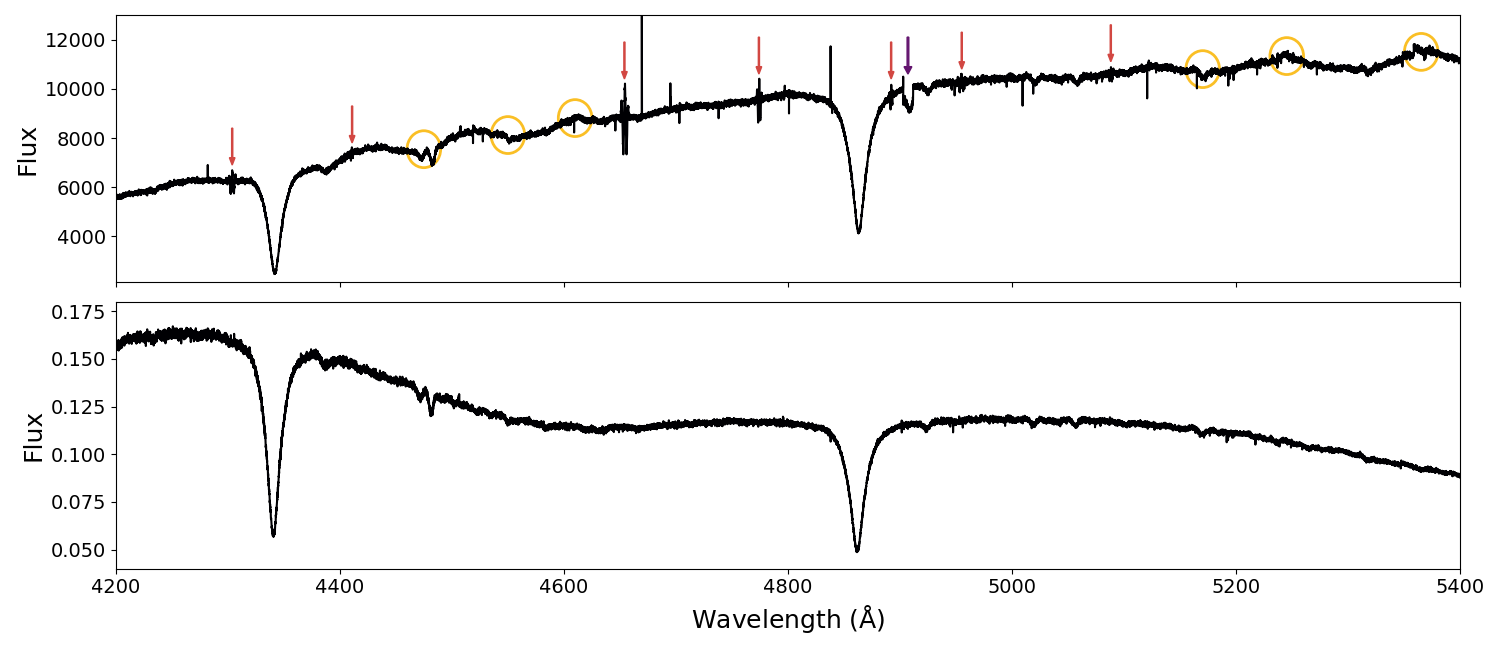}
    \caption{{\it Top:} spectrum of B9V star HD\,33244 reduced with the original CERES pipeline. Red and purple arrows indicate the positions of wiggles and a gap, yellow circles show the positions of some bumps (see the text for an explanation of these features). {\it Bottom:} spectrum of HD\,33244 reduced with the new version of the CERES pipeline that includes the four changes described in Sect.~\ref{sect:CERES}.}
    \label{fig:spectrum}
\end{figure*}

When obtaining reduced spectra with the original CERES pipeline, we detected wiggles and a gap at certain positions in the 1D spectra (see the arrows in the top panel of Fig.~\ref{fig:spectrum}). We traced these features back to bad columns in the 2D frames. The original pipeline corrects for bad columns by interpolating among the good pixels in each row at the positions of the bad pixels. However, because the orders are skewed, some parts of the bad columns are assigned too much or too little flux in this way, creating the observed wiggles in the 1D spectra. 
Instead of interpolating within each row, we found that almost all the bad columns disappear when the bias frame is subtracted from the observed science frames. The bias frame contains the same bad columns as the science frames so when the first one is subtracted from the latter, the excess in flux is removed and the bad columns are no longer visible in the 2D frames. This is the case for all but one bad column. Column 1299 is the brightest one present in the 2D frames and does not disappear by subtracting the bias. For the pixels in this column we increased the variance by 10\,000 such that they have an extremely small weight and almost no contribution in the extraction process. 
   
The original CERES pipeline corrects for the blaze function by dividing each individual order by the normalised flat field in that order. The normalised flat field is assumed to represent the blaze function according to \citet{Brahm2017}, and is defined as the flat field divided by its maximum in that order. 
However, we noticed that this blaze correction induces large intensity jumps or `bumps' where successive orders overlap (see yellow circles in the top panel of Fig.~\ref{fig:spectrum}).
When instead the orders are divided by the non-normalised flat field, they connect much better and bumps are no longer visible (see bottom panel of Fig.~\ref{fig:spectrum}). Therefore we changed the blaze correction in CERES to a division of each spectral order by the non-normalised flat field.

The CERES pipeline only outputs the 1D spectrum for individual orders. While this is fine for RV determinations, for which the pipeline was designed, we needed the whole spectrum to perform a spectral analysis. Therefore, we implemented order merging into the code. First we created a unified wavelength grid in logarithmic scale that covers the full wavelength range of all the extracted orders and has a relative step size of 10$^{-5}$. This step size is approximately equal to 1/(2$R$) with $R$ the resolution of FEROS ($R$ = 48\,000). Each order was then shifted to the new wavelength grid by linear interpolation. We merged the orders by taking a weighted sum, with as weight the square of the signal-to-noise ratio ($w$ = (S/N)$^2$):

\begin{equation} \label{eq:f_merged}
    f_{merged} = \frac{\sum_i f_i w_i}{\sum_i w_i} \quad \mathrm{and} \quad \sigma_{f_{merged}} = \frac{\sqrt{\sum_i \sigma_{f,i}^2 \ w_i^2}}{\sum_i w_i},
\end{equation}

\noindent where $f_i$ is the blaze corrected flux in order $i$ and $\sigma_{f,i}$ is its uncertainty. If pure Poisson noise is assumed, (S/N)$_i^2$ is equal to the number of photons from the star in a certain order so that the weighted mean in Eq.~\ref{eq:f_merged} is just a co-addition of photons over all orders. This means that in overlapping neighbouring orders, where part of the photons had ended up in one or the other order, all photons are added again by Eq.~\ref{eq:f_merged}.

We added two steps to the pipeline to remove sharp lines originating from cosmic rays and chip defects. In the first step we only dealt with positive hits due to cosmic hits. At every position in the spectrum we computed the median flux and standard deviation within a window of 100 pixels ($\sim$~5\AA). When the difference between the flux at that point and the median flux was larger than four times the standard deviation, the flux value was replaced with the median value. This process was repeated five times. The second step also removed negative hits due to chip defects, but here the method was more conservative to prevent removing stellar absorption lines. This is especially difficult in slowly rotating stars that have many narrow absorption lines that must not be mistaken for cosmic hits. The method is the same as for the positive cosmic hits, only the window size was changed to 30 pixels ($\sim$~1.5\AA) and both positive and negative hits were removed. Although emission lines have been detected in B-type stars \citep{Wahlgren2000,Alexeeva2016}, we are confident that our cosmic hit removal function does not remove these real emission lines since they are expected to have the same width as the absorption lines in the spectrum while only sharp and narrow features are being removed by our method.

\section{Analysis methods} \label{sect:methods}
\subsection{Spectroscopic classification} \label{sect:classification}

To make a spectroscopic classification of the sample, the reduced spectra were subjected to least-squares deconvolution \citep[LSD;][]{Donati1997}. We produced LSD profiles using the method of \citet{Tkachenko2013c}\footnote{\url{https://github.com/TVanReeth/Least-squares-deconvolution}}. 
The computation of LSD profiles requires normalised spectra and spectral line masks. Therefore, we ran each individual spectrum through the {\sc zeta-Payne} set-up, which is described in Sect.~\ref{sect:ThePayne}, and obtained normalised spectra from which the response function is removed. The line masks were computed with the GSSP code \citep{Tkachenko2015}, which relies on atmosphere models obtained with the LLmodels code \citep{Shulyak2004}, the radiative transfer code SynthV \citep{Tsymbal1996}, and Vienna Atomic Line Database \citep[VALD,][]{Kupka1999} atomic data. For each star we computed a GSSP model for its preliminary $T_{\mathrm{eff}}$ and $\log\,g$ value returned by the {\sc zeta-Payne}, and we fixed [M/H] to 0\,dex.  

\begin{figure}
    \centering
    \includegraphics[width=\columnwidth]{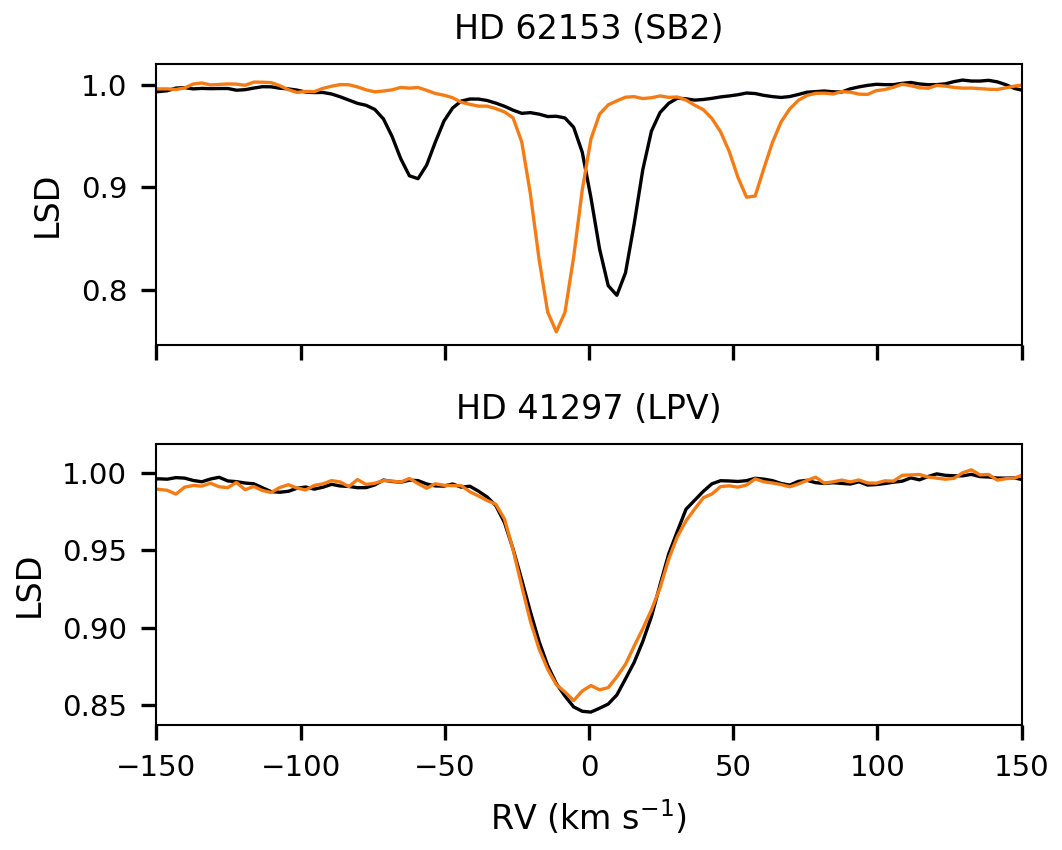}
    \caption{\textit{Top:} LSD profiles for two epochs of SB2 system HD\,62153. \textit{Bottom:} LSD profiles for two epochs of single star HD\,41297 with LPV.}
    \label{fig:LSD}
\end{figure}

We identified SB2 systems as targets for which two components could be distinguished in the LSD profiles, while in some cases also revealing different relative positions of the spectral lines of the two components for the different epochs of that target. The LSD profiles were also used to detect line profile variability (LPV), which can have various physical origins, such as chemical inhomogeneities or pulsations. Examples of LSD profiles for an SB2 system and a star with LPV are shown in Fig.~\ref{fig:LSD}.
In total 26 targets were classified as SB2 systems or as targets for which it was not clear whether the variability in the LSD profiles is due to binarity or LPV. Some of these systems were already identified as (spectroscopic) binaries in SIMBAD. These 26 targets were excluded from the sample since they require the (simultaneous) analysis of both components, which is not yet possible with our current method. They are listed as SB2 systems in Table~A.1.

The remaining targets were divided into two groups based upon the LPV in the different epochs of each target. 
The first group consists of 42 stars with LPV, either in H$\alpha$, in all Balmer lines or also in metal lines. Two of these stars are Be stars with not only Balmer lines in emission but also metal emission lines. The other 40 stars are classified as (blue) supergiants in SIMBAD and all of them are members of the LMC. For this type of star, non-local thermodynamic equilibrium (NLTE) analysis including wind physics is required to obtain atmospheric parameters. Our current set-up of the {\sc zeta-Payne} is trained on LTE model spectra for $\log\,g > 3$\,dex, while supergiants typically have $\log\,g$ values between 1 and 3\,dex. Therefore it could not be applied to the spectra of the supergiants. Instead, these 42 stars will be studied in the complementary paper by Serebriakova et al.\ (in prep.) focused on OBA-type supergiant LMC members. 

The other group contains 92 stars without clear LPV or having variability in metal lines but stable Balmer lines, as well as six Be stars with Balmer lines in emission. These 98 (Galactic) stars are further studied in this paper. 
Among them, we identified single-lined spectroscopic binaries (SB1) using the RV determinations from the {\sc zeta-Payne}. A target was classified as SB1 system when the RV difference between any two epochs of that target is larger than the uncertainties of the RV values. If the RV differences are smaller than the uncertainties or when there is only one spectrum available, the star was assumed to be single. Twelve targets were found to be SB1 systems and 86 are single stars. 
For each of the targets, we shifted all of its spectra to RV = 0\,km\,s$^{-1}$ and added them by taking a weighted sum, where the weight of each epoch is the square of the S/N value of that epoch computed following the algorithm from \citet{Stoehr2008}. We subsequently used that same algorithm to compute the S/N of the averaged spectrum; these values are reported in column 3 of Table~A.2. The number of epochs and binary information are given in columns 2 and 4 of the same table.

\begin{figure}
    \centering
    \includegraphics[width=\columnwidth]{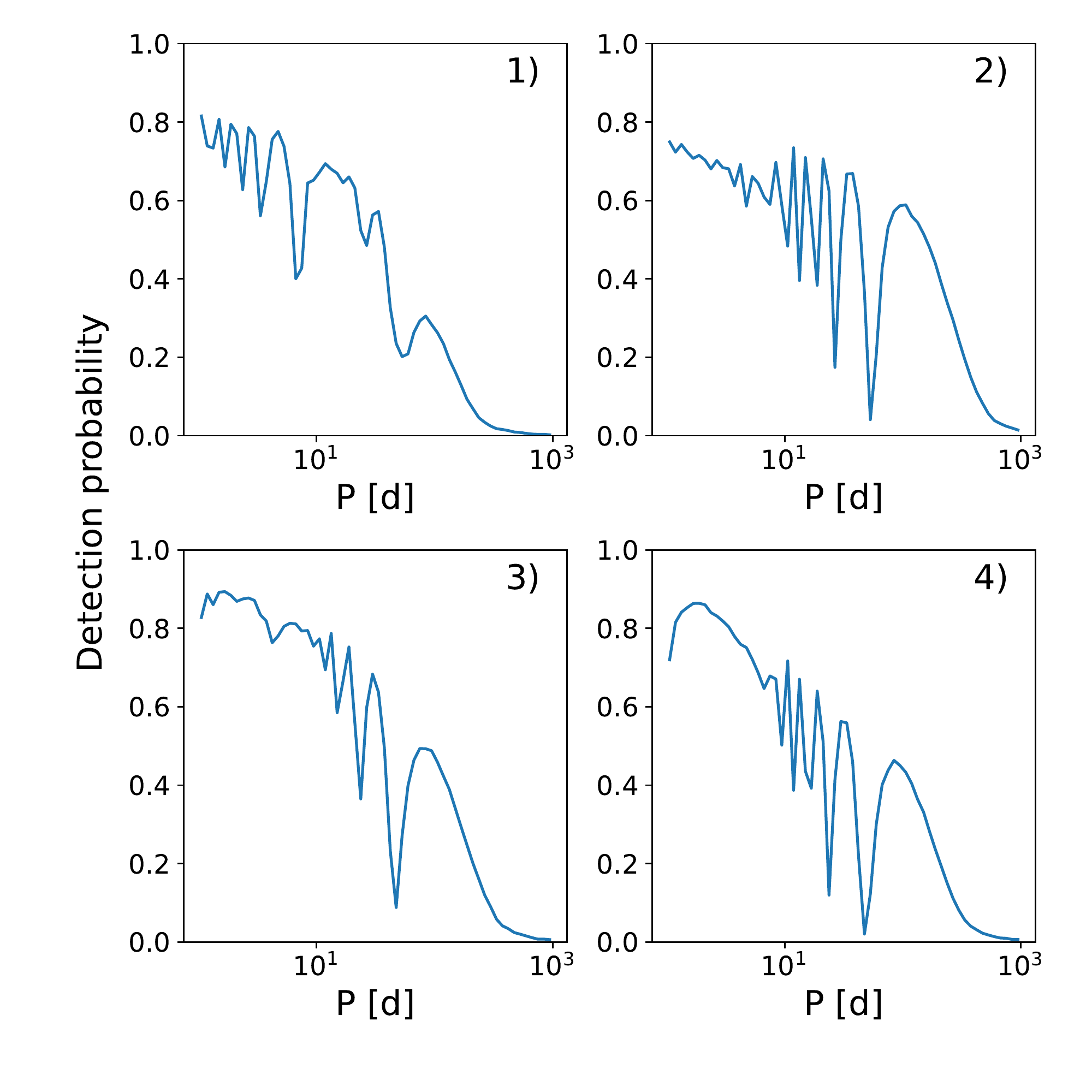}
    \caption{Binary detection probabilities as a function of the orbital period for four different observing strategies used to constitute our dataset: 1) two observations were obtained within one week and one observation one month and a half later, 2) two observations were obtained spread over $\sim 2$ months, 3) four observations were obtained: two during one week and two collected during another week two months later and 4) two observations were obtained on consecutive nights and one observation two months later.}
    \label{fig:detectionProb}
\end{figure}

We computed the binary detection probability for all our targets. We distinguished four different observing strategies to obtain the observations required for our study: 1) two observations were obtained within one week and one observation one month and a half later, 2) two observations were obtained spread over two months, 3) four observations were obtained: two during one week and two collected during another week two months later and 4) two observations were obtained on consecutive nights and one observation two months later. Fig.~\ref{fig:detectionProb} displays the binary detection probabilities in these different campaigns. To compute these probabilities, we ran 1 million Monte Carlo simulations tuned for each observing strategy. We took a uniform period distribution in logarithmic space ranging from 1 to 1000 days and eccentricity values between 0 and 0.95, with all orbits with eccentricities lower than 0.03 considered as circular. The mass-ratio distribution was considered uniform from 0.01 to 1 and the initial mass for each star was taken between 2 and 5~M$_{\odot}$, that is, in agreement with their spectral classification \citep{Schmidt-Kaler1982}. We adopted a Salpeter initial mass function (IMF, with $\alpha = -2.35$) for the primary masses. Our simulations assumed that the orbits are randomly oriented in the three-dimensional space, the time of periastron passage is uncorrelated with respect to the start of the RV campaign, and that the orbital parameters are uncorrelated. To detect a system as binary, it needs to fulfil the same criteria as the ones we adopted for our analysis.

For orbital periods shorter than 10 days, the probability to detect binary systems is rather high with a probability of 60\% or higher, whatever the observing strategy used. For systems with periods between 10 and 100 days, we still reach a detection rate between 30 and 60\%. This rate significantly drops for longer-period systems, and it is almost not possible to detect binary systems with orbital periods larger than 1000 days.

\subsection{Spectrum analysis with the {\sc zeta-Payne}} \label{sect:ThePayne}

We used the {\sc zeta-Payne} \citep{Straumit2022} spectrum analysis algorithm to analyse the averaged spectra of the 98 Galactic single stars or SB1 systems. The {\sc zeta-Payne} is a machine learning framework that trains a neural network on a grid of synthetic spectra to obtain a spectrum interpolator. This interpolator predicts theoretical flux spectra for an arbitrary set of stellar labels and can be used to fit any observed spectrum and derive its surface parameters as long as they are within the parameter ranges of the training sample. 

In this paper, the neural network was trained with 1D LTE synthetic models computed with GSSP. Our choice of the LTE formalism is well justified given that the $T_{\mathrm{eff}}$ distribution of the sample stars is found to peak at some 11\,500\,K, with only one target being hotter than 20\,000\,K (cf. Fig.~\ref{fig:histograms}). In their recent study of OBA-type stars in the LAMOST survey, \citet{Xiang2021} demonstrate the impact of NLTE effects on the spectroscopically inferred $T_{\mathrm{eff}}$ values to be negligible for stars cooler than 25\,000\,K. In particular, the authors show an overall good consistency between the Kurucz LTE and Potsdam Wolf-Rayet \citep[PoWR][]{Hainich2019} NLTE models at $T_{\mathrm{eff}}\lessapprox$ 25\,000\,K (their Figs.~C.1 and C.2), and report a good agreement between the parameters obtained with the LTE and NLTE models for stars below that same $T_{\mathrm{eff}}$ cut-off value (their Fig.~11). Instead of using a regular grid in five dimensions -- one for each of the parameters we want to determine, namely $T_{\mathrm{eff}}$, $\log\,g$, global metallicity ([M/H]), projected rotational velocity ($v\sin\,i$) and microturbulent velocity ($\xi$) -- we created a grid that is quasi-randomly spaced using Sobol numbers \citep{Sobol1967}. It covers the whole parameter space for main sequence BAF-type stars as listed in the second column of Table~\ref{tab:training_grid}, but with fewer grid points than a regular grid would need and is therefore more feasible to compute.
\citet{Straumit2022} point out that, for high-resolution spectra, the neural network performs worse in regions of the parameter space with low values of $v\sin\,i$ and temperatures in the late-A to F-type regime. This is due to the many narrow spectral lines that have to be resolved by the neural network in this regime. To improve the performance of the neural network for such complex spectra, the parameter space should be oversampled with model spectra computed for low $v\sin\,i$ and low $T_{\mathrm{eff}}$ values.   
The final training sample therefore consists of 5000 quasi-randomly spaced model spectra and 5000 additional random model spectra with $T_{\mathrm{eff}}$ and $v\sin\,i$ drawn from the normal distributions $\mathcal{N}$(6000, 4000) and $\mathcal{N}$(0, 25) respectively. An overview of the parameter ranges for the whole training grid is given in Table~\ref{tab:training_grid}. 

\begin{table}
    \centering
    \caption{Parameter ranges of the training grid.}
    \begin{tabular}{lcc}
        \hline \hline
        Parameter & Quasi-random grid & Oversampled grid \\
        \hline 
        $T_{\mathrm{eff}}$ (K) & 6000 - 25\,000 & $\mathcal{N}$(6000, 4000) \\
        $\log\,g$ (dex) & 3 - 5 & 3 - 5 \\
        $v\sin\,i$ (km\,s$^{-1}$) & 0 - 400 & $\mathcal{N}$(0, 25) \\
        $\xi$ (km\,s$^{-1}$) & 0 - 20 & 0 - 20\\
        $\mathrm{[M/H]}$ (dex) & -0.8 - +0.8 & -0.8 - +0.8 \\
        \hline
    \end{tabular}
    \label{tab:training_grid}
\end{table}

All the model spectra were computed within the wavelength range from 3000 to 10\,500 \AA \ and for infinite resolution (no instrumental broadening) such that the grid can also be used for optical spectra from spectrographs other than FEROS.   
During the fitting procedure the spectra are convolved to the resolution of the respective spectrograph by either using a constant resolution value or a wavelength dependent resolution (the line spread function, LSF). We used the constant resolution value of FEROS, because the LSF computed from the ThAr frames that were taken during the observing run, changes from night to night.
The training of the neural network was done identical to the procedure described in \citet{Straumit2022}, so more information can be found in that paper. 
For every star in the sample, surface parameters and the RV were derived by fitting model spectra to the observed spectrum using the neural network interpolator, performing a Doppler shift, and minimising the $\chi^2$ merit function. However, the neural network was trained on continuum normalised spectra while the observed spectra contain components from the instrument (the response function), the interstellar medium (ISM) and the Earth's atmosphere. Each observed spectrum had to be divided by its response function before it could be compared to the normalised model spectra. In the ideal case, the response function of a spectrograph is known and can be removed from any observed spectrum to obtain only the component from the star itself (and some additional features from the ISM and the Earth's atmosphere such as telluric lines). This is attempted by dividing an observed spectrum by a flat-field exposure, but some curvature always remains in the spectrum that is not related to the stellar atmosphere \citep[][Chap. 12]{Gray2005}. In practice, the response function is mostly removed by fitting a spline or a polynomial to manually selected points that are assumed to be part of the continuum. This is a rather subjective way of normalising a spectrum and can give different results when it is done by different researchers.

In order to minimise subjectivity, we followed \citet{Straumit2022} and normalised the spectra automatically during the fitting procedure in the {\sc zeta-Payne}, such that the optimal response function was determined simultaneously with the best-fitting surface parameters. We assumed that the response function could be represented by a Chebyshev polynomial for which we optimised its coefficients \citep[as described in][]{Straumit2022}. The number of coefficients is a free parameter that had to be fixed before starting the fitting procedure. Preferably the number is as small as possible to ensure that no features are introduced into the spectrum, but it should also be high enough to capture the whole shape of the response function.
It also depends on the wavelength range, where generally more coefficients are needed when a longer wavelength range is used. For our analysis, we selected a wavelength range from 4000 to 5800 \AA\ because it includes three Balmer lines (H$\beta$, H$\gamma$ and H$\delta$) and many metal lines, but excludes the H$\alpha$ region that complicates the optimal normalisation of FEROS spectra.
Since all the response functions of the FEROS spectra have a similar shape, the same number of Chebyshev coefficients could be used for all the stars. 

\begin{figure}
    \centering
    \includegraphics[width=0.9\columnwidth]{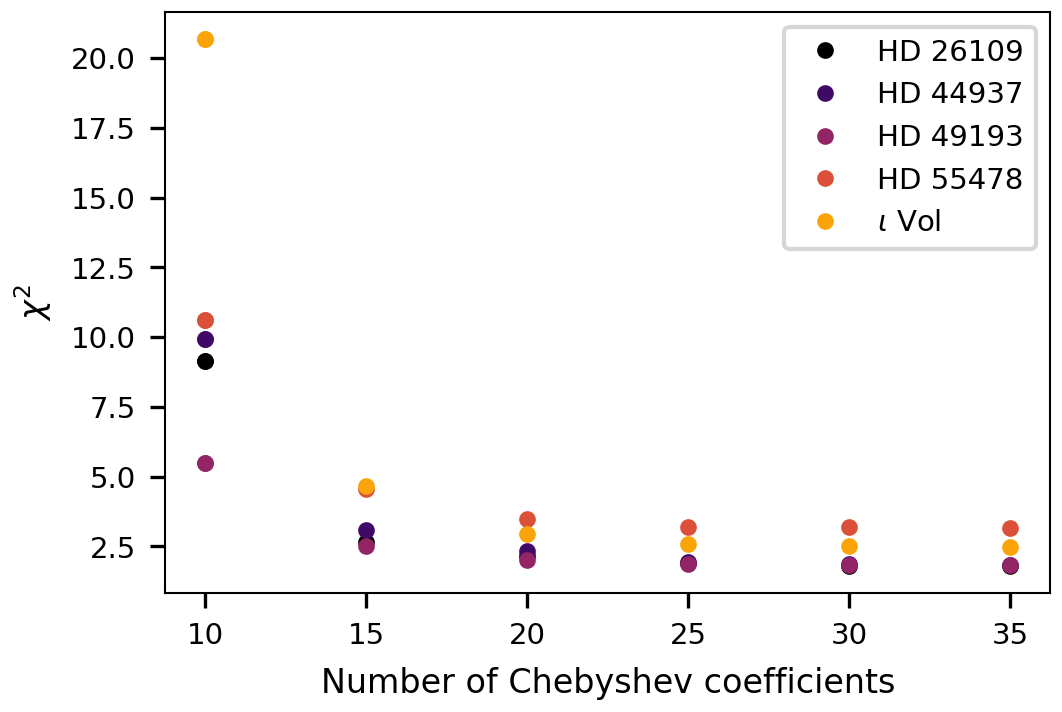}
    \caption{$\chi^2$ values for best-fitting synthetic spectra to the observed spectra of five sample stars as a function of the number of Chebyshev coefficients.}
    \label{fig:num_coeffs}
\end{figure}

We tested the optimal number of Chebyshev coefficients for a typical FEROS response function by taking five random stars from the sample for which we fitted the spectra with 10, 15, 20, 25, 30 and 35 coefficients with the {\sc zeta-Payne}. For each fit we computed the $\chi^2$ value and plotted those values as a function of the number of coefficients (see Fig.~\ref{fig:num_coeffs}). The $\chi^2$ profile converges towards a certain value and this plateau is reached around 25 coefficients. Therefore, we used 25 Chebyshev coefficients to represent the FEROS response functions. 

\subsection{Internal uncertainties} \label{sect:sys_unc}
We determined the internal uncertainties of the stellar parameters that are inherent to the modelling set-up. We created 1000 artificial FEROS spectra in the wavelength range from 4000 to 5800~\AA, quasi-randomly spaced within the parameter space but not overlapping with the training sample. Each spectrum was computed with GSSP, given a random RV shift within $\pm 50\,\mathrm{km\,s}^{-1}$, and convolved to the resolution of FEROS. A random response function was introduced represented by a Chebyshev polynomial with 25 coefficients. For every spectrum we got a noiseless version and a version with S/N = 150, a typical value of the real FEROS spectra. All the artificial spectra were analysed with the fitting routine of the {\sc zeta-Payne} and the resulting parameters were compared with their real values. 
Outliers were identified as spectra for which the $T_{\mathrm{eff}}$ difference was larger than four times the 1$\sigma$ standard deviation of the $T_{\mathrm{eff}}$ difference distribution of all 1000 spectra. This is formulated as the reliability fraction of the neural network \citep{Straumit2022} and is given in the last row of Table~\ref{tab:sys_uncert_all}.

Using only the remaining spectra without the outliers, we computed the internal uncertainty for each parameter as the mean difference between the real values and the ones obtained with the {\sc zeta-Payne}. These uncertainties for noiseless spectra and for spectra with S/N = 150 are given in Table~\ref{tab:sys_uncert_all}, and should be taken into account when determining parameters for the observed FEROS spectra.   

\begin{table}
    \centering
    \caption{Average internal uncertainties for the surface parameters and the RV computed from 1000 simulated FEROS spectra, without noise and for S/N = 150.}   
    \begin{tabular}{lcc}
    \hline\hline
    & \multicolumn{2}{c}{Internal uncertainty} \\
    Parameter & Noiseless & S/N = 150 \\
    \hline
    $T_{\mathrm{eff}}$ (K) & 430 & 425 \\
    $\log\,g$ (dex) & 0.12 & 0.12 \\
    $v\sin\,i$ (km\,s$^{-1}$) & 12 & 12 \\
    $\xi$ (km\,s$^{-1}$) & 2.20 & 2.25 \\
    $\mathrm{[M/H]}$ (dex) & 0.13 & 0.13 \\
    RV (km\,s$^{-1}$) & 0.96 & 0.97 \\
    \hline
    Reliability & 94.6\% & 94.3\% \\
    \hline
    \end{tabular}
    \label{tab:sys_uncert_all}
\end{table}

\begin{figure*}
    \centering
    \includegraphics[width=0.9\textwidth]{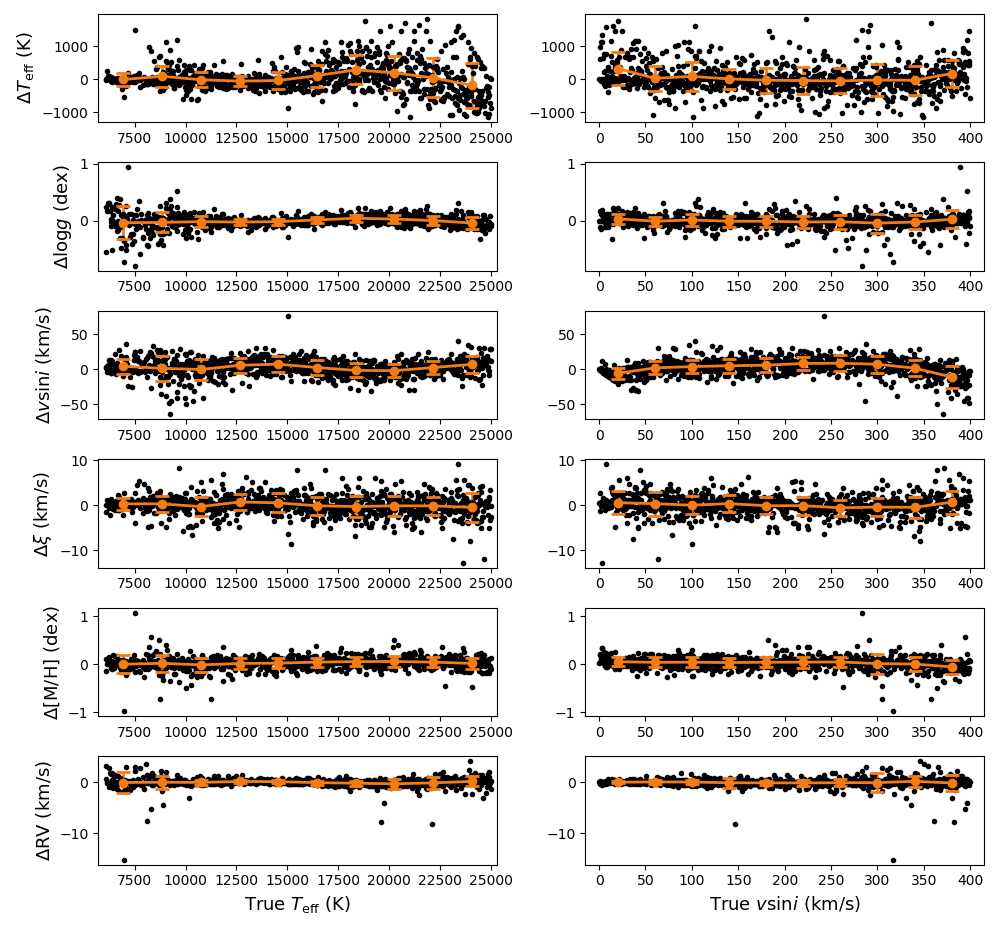}
    \caption{Internal uncertainties (the difference between parameters predicted with the {\sc zeta-Payne} and the true values) for 1000 synthetic noiseless FEROS spectra shown as a function of $T_{\mathrm{eff}}$ (\textit{left}) and $v\sin\,i$ (\textit{right}). The orange dots are the average uncertainties with standard deviation in bins of $T_{\mathrm{eff}}$ and $v\sin\,i$.}
    \label{fig:sys_uncert}
\end{figure*}

Apart from the total internal uncertainties for the full $T_{\mathrm{eff}}$ range of the grid, we also computed the internal uncertainties for the noiseless spectra in bins of $T_{\mathrm{eff}}$ and $v\sin\,i$. An overview of this can be found in Table~\ref{tab:sys_uncert_bin} and is plotted in Fig.~\ref{fig:sys_uncert}. 
One of the interesting features in this figure is the tail of high $T_{\mathrm{eff}}$ uncertainties in the $\sim$\,9000\,K regime. A closer look at the properties of spectra forming the tail reveals low [M/H] and high $v\sin i$ values which in turn implies apparent dearth of spectral lines of metals. This significant loss of information, where one has to rely exclusively on the Balmer lines to estimate the $T_{\mathrm{eff}}$-$\log\,g$ pair of parameters, results in larger uncertainties in $T_{\mathrm{eff}}$ and, to a lesser extent, $\log\,g$ of the star. There is also an increase in absolute uncertainty towards high $T_{\mathrm{eff}}$ values, yet relative uncertainties remain at the level of 2-3\%, similar to the lower $T_{\mathrm{eff}}$ regime (see Table~\ref{tab:sys_uncert_bin}).
We also note scatter in $\log\,g$ for spectra with $T_{\mathrm{eff}} < 10\,000$K, that is for late A- and F-type stars. As discussed in \citet{Gebruers2021}, the $\log\,g$ determination for these stars is degenerate with $T_{\mathrm{eff}}$, [M/H] and the continuum normalisation of the spectrum. Small offsets in the continuum result in large differences in $\log\,g$, which explains the larger internal uncertainty in this $T_{\mathrm{eff}}$ bin. 
The same trends in the internal uncertainties are reported by \citet{Straumit2022} for APOGEE (infrared) and BOSS (optical) spectra. 

\begin{table*}[ht]
    \centering
    \small
    \caption{Internal uncertainties for the surface parameters and RV in bins of $T_{\mathrm{eff}}$ and $v\sin\,i$ computed from a sample of 1000 simulated noiseless FEROS spectra.}    
    \begin{tabular}{lllllllllll}
    \hline\hline
    & \multicolumn{10}{c}{Internal uncertainty}  \\
    Parameter & Bin 1 & Bin 2 & Bin 3 & Bin 4 & Bin 5 & Bin 6 & Bin 7 & Bin 8 & Bin 9 & Bin 10 \\
    \hline
    & \multicolumn{10}{c}{As function of $T_{\mathrm{eff}}$} \\
    \hline
    Bin (kK) & 6-7.9 & 7.9-9.8 & 9.8-11.7 & 11.7-13.6 & 13.6-15.5 & 15.5-17.4 & 17.4-19.3 & 19.3-21.2 & 21.2-23.1 & 23.1-25 \\
    $T_{\mathrm{eff}}$ (K) & 211 (3\%) & 321 (4\%) & 226 (2\%) & 161 (1\%) & 267 (2\%) & 334 (2\%) & 438 (2\%) & 515 (3\%) & 597 (3\%) & 679 (3\%) \\
    $\log\,g$ (dex) & 0.29 & 0.17 & 0.10 & 0.05 & 0.05 & 0.05 & 0.06 & 0.07 & 0.08 & 0.10 \\
    $v\sin\,i$ (km\,s$^{-1}$) & 11 & 18 & 14 & 10 & 11 & 10 & 10 & 9 & 9 & 12 \\
    $\xi$ (km\,s$^{-1}$) & 1.3 & 1.9 & 2.1 & 1.9 & 2.1 & 2.0 & 2.3 & 2.2 & 2.1 & 3.3 \\
    $\mathrm{[M/H]}$ (dex) & 0.19 & 0.18 & 0.15 & 0.11 & 0.10 & 0.09 & 0.10 & 0.11 & 0.11 & 0.12 \\
    RV (km\,s$^{-1}$) & 2.1 & 1.3 & 0.5 & 0.3 & 0.3 & 0.4 & 0.5 & 1.0 & 1.1 & 1.0 \\
    \hline
    & \multicolumn{10}{c}{As function of $v\sin\,i$} \\
    \hline
    Bin (km\,s$^{-1}$) & 0-40 & 40-80 & 80-120 & 120-160 & 160-200 & 200-240 & 240-280 & 280-320 & 320-360 & 360-400 \\
    $T_{\mathrm{eff}}$ (K) & 509 & 374 & 433 & 315 & 380 & 402 & 378 & 479 & 434 & 418 \\
    $\log\,g$ (dex) & 0.09 & 0.08 & 0.10 & 0.10 & 0.09 & 0.11 & 0.13 & 0.17 & 0.13 & 0.16 \\
    $v\sin\,i$ (km\,s$^{-1}$) & 8 & 9 & 9 & 10 & 10 & 9 & 12 & 12 & 11 & 15 \\
    $\xi$ (km\,s$^{-1}$) & 2.5 & 2.7 & 2.0 & 1.8 & 1.9 & 2.0 & 1.7 & 2.0 & 2.3 & 2.6 \\
    $\mathrm{[M/H]}$ (dex) & 0.09 & 0.10 & 0.10 & 0.09 & 0.11 & 0.11 & 0.11 & 0.21 & 0.14 & 0.15 \\
    RV (km\,s$^{-1}$) & 0.2 & 0.4 & 0.5 & 1.0 & 0.4 & 0.5 & 0.7 & 1.8 & 1.1 & 1.6  \\
    \hline
    \end{tabular}
    \label{tab:sys_uncert_bin}
\end{table*}

\subsection{Effect of the selected wavelength region}\label{sec:wave_region}

\begin{figure*}
    \centering
    \includegraphics[width=\textwidth]{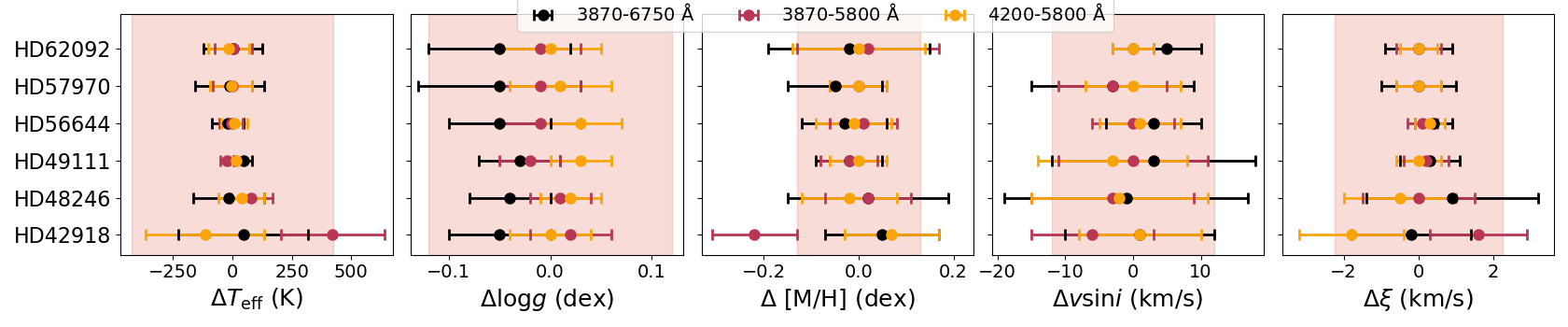}
    \caption{Comparison of stellar parameters derived from different wavelength regions. From \textit{left} to \textit{right} the differences in $T_{\mathrm{eff}}$, $\log\,g$, [M/H], $v\sin\,i$ and $\xi$ are shown. Black, red and yellow dots are the differences between the parameter value obtained from the 4000-5800\,\AA \ spectral range and the parameter value obtained from the wavelength range indicated with the respective colour in the legend of the figure, so $p_{4000-5800\mathrm{\AA}}$ - $p_{\Delta\lambda}$. The pink region indicates the internal uncertainty interval from Table~\ref{tab:sys_uncert_all}. The systematic (though insignificant within the internal uncertainty) offset in the $\log\,g$ parameter is discussed in detail in Section~\ref{sec:wave_region}.}
    \label{fig:test_diff_wavelength}
\end{figure*}

The atmospheric parameters of the main sequence stars in our sample were derived using the spectral wavelength range 4000-5800\,\AA. To test whether the analysis of a longer or shorter wavelength range has any effect on the obtained parameters, we selected six stars distributed over the parameter space and applied the {\sc zeta-Payne} on three different wavelength ranges. The wavelength ranges are: 3870-6750\,\AA \ (H$\alpha$, H$\beta$, H$\gamma$, H$\delta$, and H$\epsilon$), 3870-5800\,\AA \ (H$\beta$, H$\gamma$, H$\delta$, and H$\epsilon$), and 4200-5800\,\AA \ (H$\beta$ and H$\gamma$). 
For each of these wavelength ranges, we computed the difference in parameters with those obtained from the original wavelength range (4000-5800\,\AA). This is shown in Fig.~\ref{fig:test_diff_wavelength}. All the values lie within the internal uncertainties from Table~\ref{tab:sys_uncert_all}. This demonstrates that the choice of wavelength region has no statistically significant effect on the surface parameters when the internal uncertainties are taken into account. 

\begin{figure*}
    \centering
    \includegraphics[width=\textwidth]{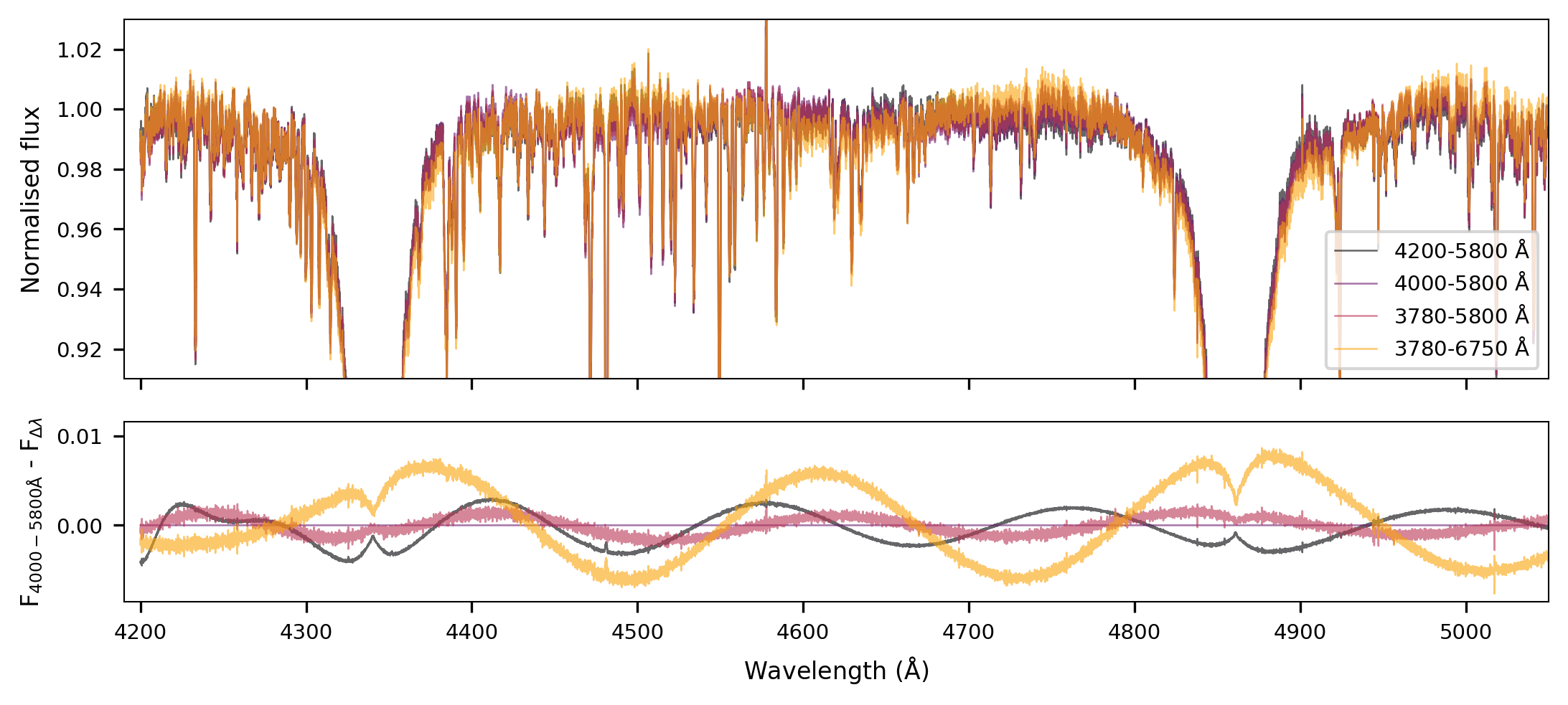}
    \caption{Comparison of the normalisation quality when different wavelength regions are analysed. \textit{Top panel:} spectra computed in the 4200-5800\,\AA \ (black), 4000-5800\,\AA \ (purple), 3780-5800\,\AA \ (red), 3780-6750\,\AA \ (yellow) wavelength regions are shown. \textit{Bottom panel} shows the differences between the spectrum computed in the 4000-5800\,\AA \ range and those computed in the other wavelength regions as indicated by the same colours as in the top panel.}
    \label{fig:logg_wavelength}
\end{figure*}

We noticed that the uncertainties on the surface parameters increase with longer wavelength regions. There is also a trend of increasing $\log\,g$ when a longer wavelength region is analysed, although all values are still within the internal uncertainties. This is caused by the normalisation of the spectrum, which becomes more difficult when more parts of the spectrum are included. Especially when H$\alpha$ is included, a wavelength range containing multiple telluric lines is covered and the {\sc zeta-Payne} cannot model these.
For the stars in Fig.~\ref{fig:test_diff_wavelength}, we found that for longer wavelength regions, the wings of the Balmer lines get more broadened than when shorter wavelength regions are analysed with the {\sc zeta-Payne}. This is demonstrated in Fig.~\ref{fig:logg_wavelength} and shows that our normalisation approach with a Chebyshev polynomial of 25 coefficients is not optimised to normalise a spectrum extended to H$\alpha$ or to H$\epsilon$, and that the deterioration in normalisation of the Balmer wings gets compensated by preferring a model with higher $\log\,g$. Additional tests for the number of Chebyshev coefficients and methods to remove tellurics are needed to get a better normalisation in each specific wavelength range. Here we decided to limit our analysis to the 4000-5800\,\AA \ region since not that much information is added by extending to H$\epsilon$ and normalisation becomes much more uncertain when including H$\alpha$. 

\section{Analysis results} \label{sect:results}
\subsection{Spectroscopy}

\begin{figure*}
    \centering
    \includegraphics[width=0.35\textwidth]{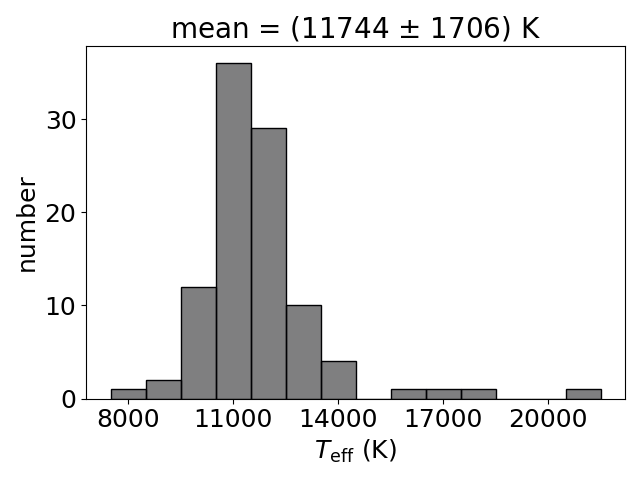}
    \includegraphics[width=0.35\textwidth]{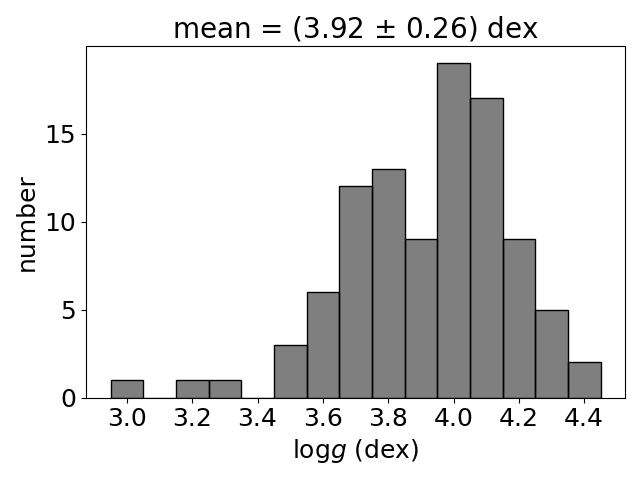}
    \includegraphics[width=0.35\textwidth]{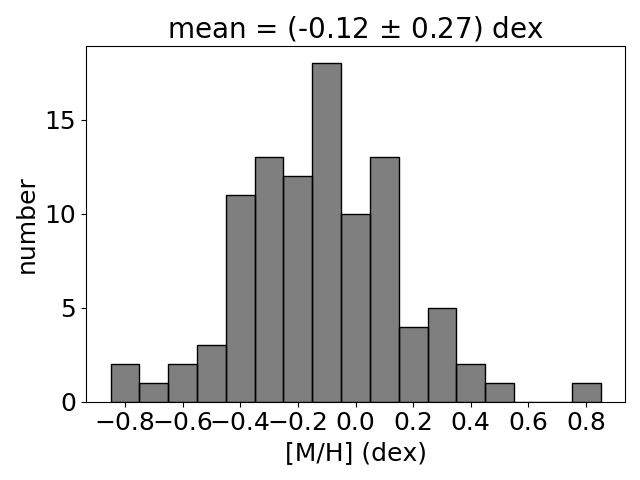}
    \includegraphics[width=0.35\textwidth]{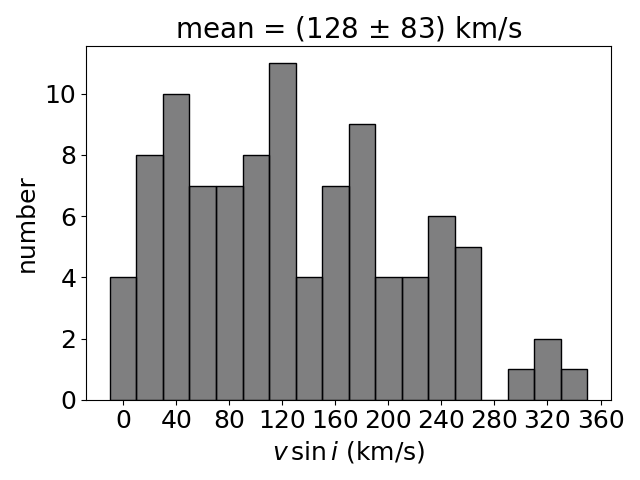}
    \caption{Distributions of $T_{\mathrm{eff}}$, $\log\,g$, [M/H] and $v\sin\,i$ for the sample of single and SB1 Galactic stars.}
    \label{fig:histograms}
\end{figure*}

The results for the 98 stars analysed with {\sc zeta-Payne} are given in Table~A.2. The statistical uncertainties inferred from using {\sc zeta-Payne} ($\sigma$) and those that include the internal uncertainty ($\sigma_{\mathrm{i}}$) are also provided. We advise to use the latter value since they are more representative for the typical precision with which the surface parameters of B-type stars can be derived. 
The distributions of the surface parameters are shown in Fig.~\ref{fig:histograms}.
Most of the stars have parameters indicative of late B-type, with temperatures in the range 10\,000-14\,000\,K. There are some cooler stars (early A-type) in the sample and a few hotter stars with $T_{\mathrm{eff}} > 16\,000$\,K. From the $\log\,g$ distribution it can be seen that the majority are main-sequence stars. However, there are six stars with $\log\,g <$ 3.5\,dex indicating that these are near the end or just beyond the main sequence. The $v\sin\,i$ values are fairly uniformly distributed within 0-350\,km\,s$^{-1}$ and [M/H] shows a normal distribution around solar value with mean of ($-0.12 \pm 0.27$)\,dex. We note that the metallicity [M/H] refers in our case to all chemical elements heavier than helium.

All the fits with the {\sc zeta-Payne} were visually inspected. For 20 stars we noticed that the model predicted by the {\sc zeta-Payne} as best fitting model does not fully correspond to the observed spectrum. 
In 13 of those cases, the observed spectrum contains more absorption lines than the predicted spectrum. These stars were identified as chemically peculiar, and specifically HgMn stars, which are further discussed in Sect.~\ref{sect:HgMn}. 
Three stars (HD\,40435, HD\,45527, and HD\,63928) show strong LPV in their spectra and four other ones (CPD-60\,944B, HD\,32493, HD\,59426, and HD\,6783) have very narrow spectral lines with strongly variable depths between different epochs. These stars are indicated as `LPV' and `LPV*', respectively, in Table~A.2. From the TESS light curves discussed in the following section, we find that these seven stars are rotational variables. This means that they are likely chemically peculiar (that is, ApBp stars), which explains the limited agreement with the best fitting model spectra. In many of these stars Cr is overabundant, confirming the chemically peculiar nature. An in-depth chemical abundance analysis is beyond the scope of this work.

\subsection{Variability classification from TESS Photometry} \label{section: photometry}

We analysed both SAP and PDC-SAP light curves independently for all stars to gauge the possible effect of systematics, and we checked that the allocated aperture mask and background pixels by the SPOC pipeline are reasonable. It has been shown that, for example, SAP light curves are preferable in the identification and analysis of bright eclipsing binaries (see e.g. \citealt{Southworth2020c, Southworth2021a}). For each star, we calculated the amplitude spectrum of its SAP and PDC-SAP light curves using a discrete Fourier transform (DFT; \citealt{Kurtz1985b}) up to the Nyquist frequency, which for the 2-min TESS data is 360\,d$^{-1}$. A high diversity and incidence of photometric variability occurs in massive stars, with common causes including pulsations, rotational modulation, winds and binarity \citep{Pedersen2019, Bowman2019a, Bowman2019b, Bowman2020b, Burssens2020}. Based on a group of expert classifiers, we visually inspected the light curves and amplitude spectra and determined the dominant variability type(s), which is included in Table~A.1. 

Of particular interest are stars pulsating in coherent p and/or g modes for follow-up asteroseismic modelling (e.g. \citealt{Moravveji2015,Moravveji2016,Szewczuk2018,Szewczuk2022,Pedersen2021}), (pulsating) eclipsing binaries (e.g. \citealt{Tkachenko2020a, Sekaran2020, Sekaran2021}) and stars with SLF variability caused by gravity waves (e.g. \citealt{Bowman2019b, Bowman2020b}). A full list of all variability types used in this work, which are not mutually exclusive, include:
\begin{itemize}
    \item rot: rotational modulation;
    \item SPB: slowly pulsating B star (i.e. predominantly g-mode pulsations);
    \item const: constant star (i.e. no significant variability);
    \item EB: eclipsing binary;
    \item EV: ellipsoidal variable;
    \item $\beta$\,Cep: early-B star (i.e. predominantly p-mode pulsations);
    \item $\delta$\,Sct: A- and early-F star with p- and g-mode pulsations;
    \item SLF: stochastic low-frequency variability;
\end{itemize}
\noindent Stars for which the variability type is also assigned a question mark (?) have a tentative classification. Also, in some cases, the inferred $T_{\mathrm{eff}}$ and $\log\,g$ values from our spectroscopic analysis were used to separate, for example, $\delta$\,Sct and $\beta$\,Cep pulsators, which can have similar amplitude spectra.

\begin{figure}
    \centering
    \includegraphics[width=\columnwidth]{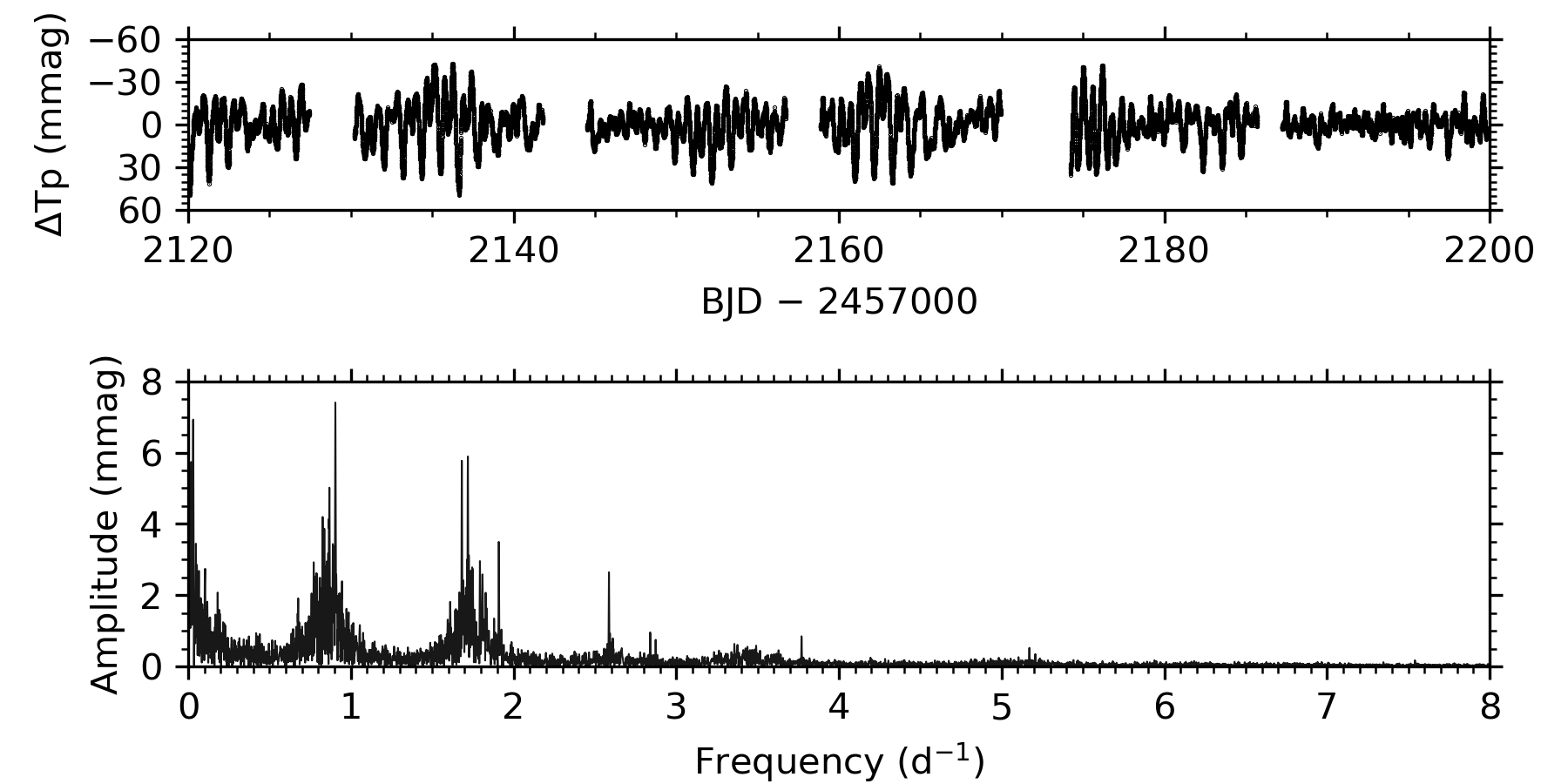}
    \includegraphics[width=\columnwidth]{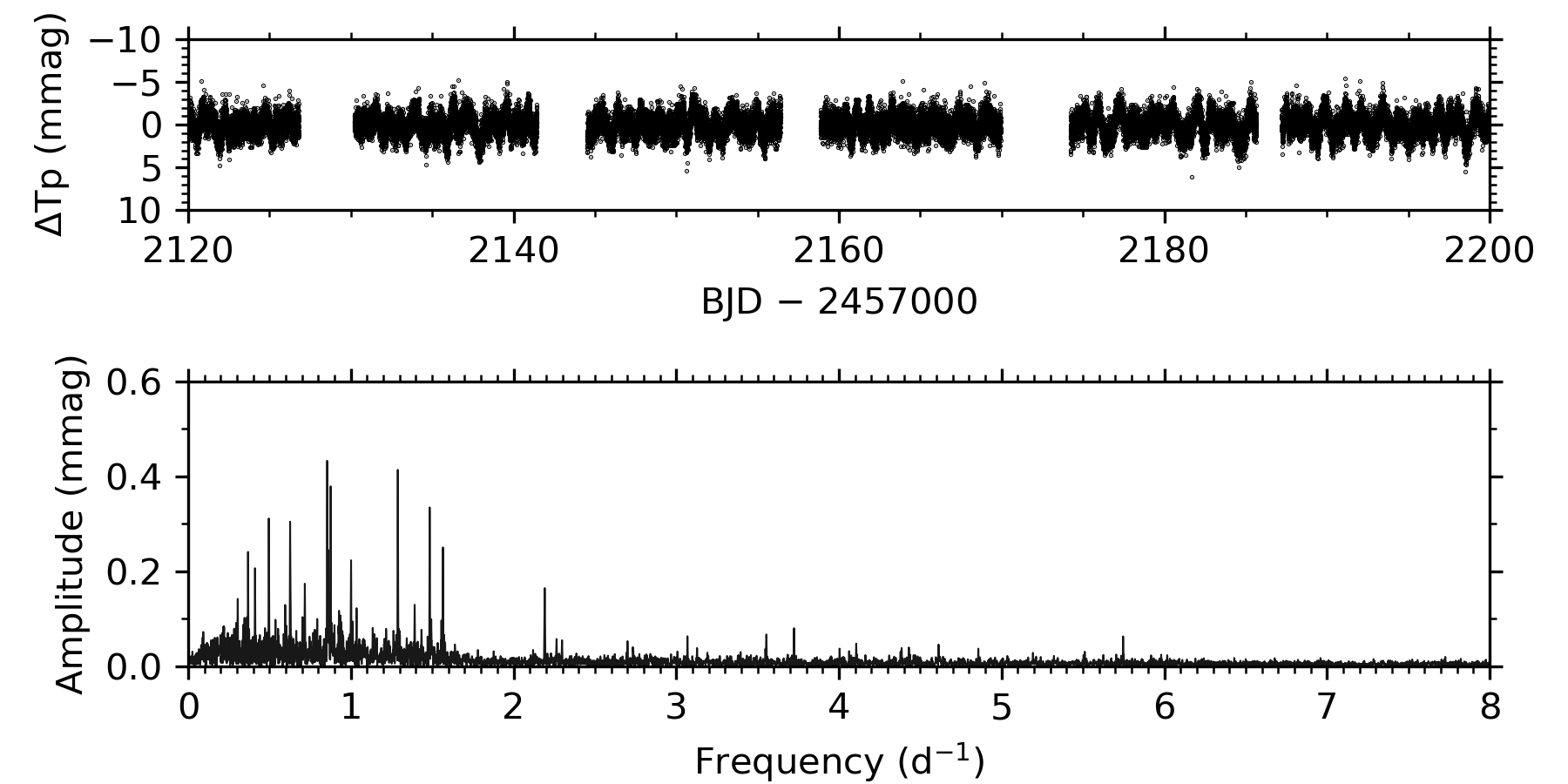}
    \includegraphics[width=\columnwidth]{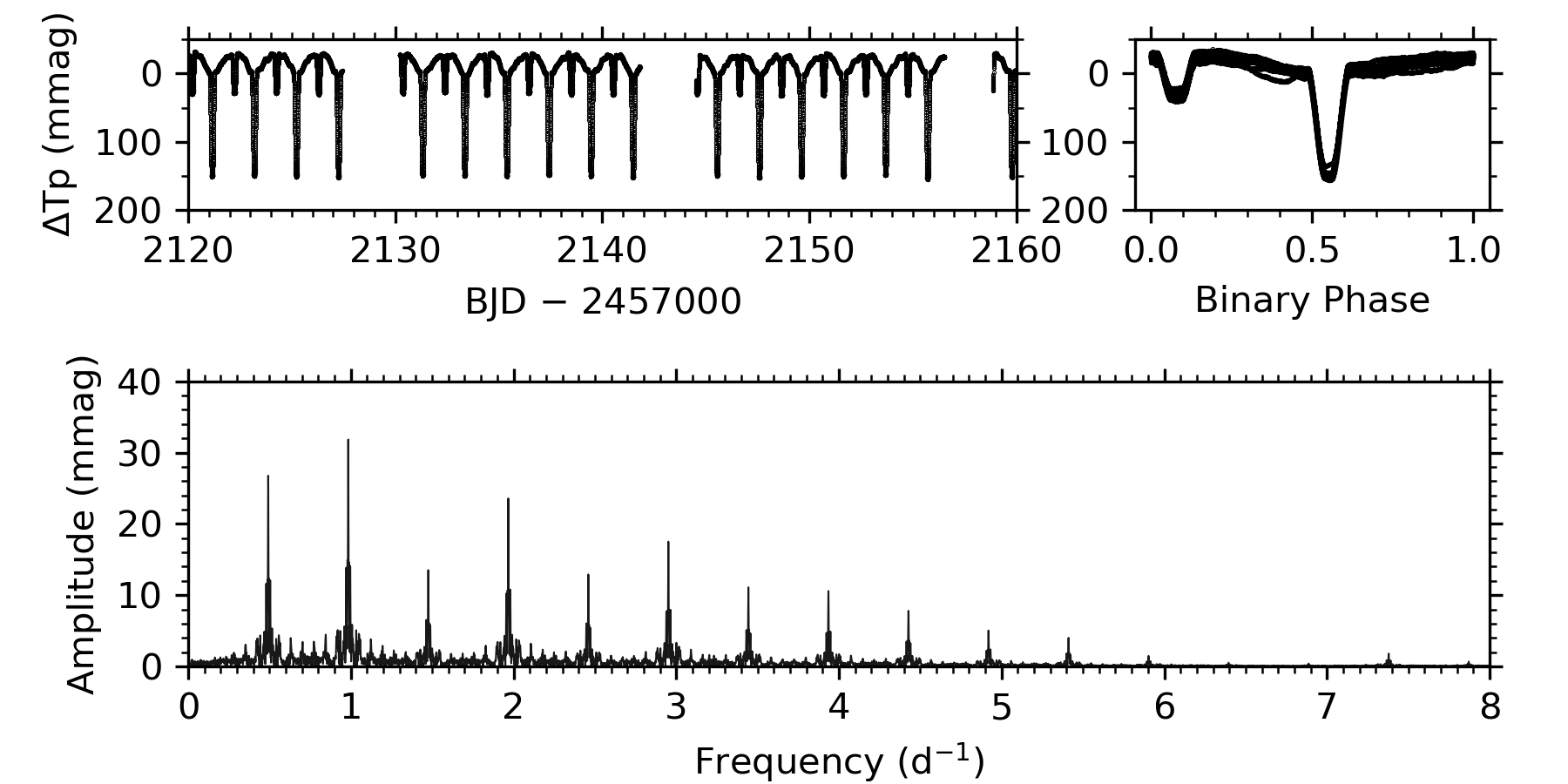}
    \includegraphics[width=\columnwidth]{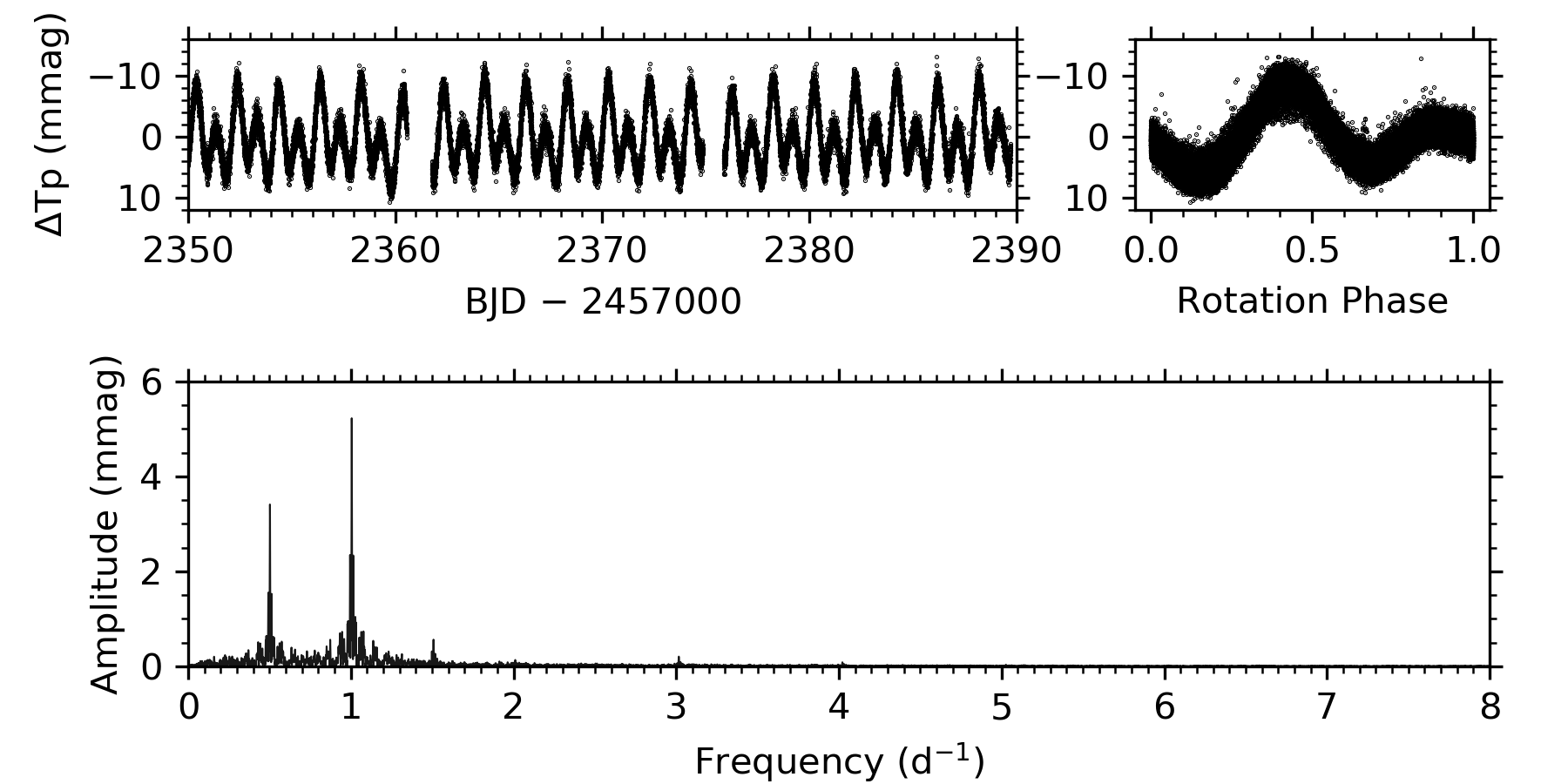}
    \caption{Subsection of the 2-min TESS light curve and amplitude spectrum of four example stars. From \textit{top} to \textit{bottom}: TIC\,55295028 (HD\,33599; SPB), TIC\,167523976 (HD\,49193; SPB / $\beta$\,Cep), TIC\,220430912 (AN\,Dor; EB), and TIC\,31313111 (HD\,40435; rot). For the latter two stars, a phase-folded light curve is shown in the top-right panel.}
    \label{fig:lightcurves}
\end{figure}

We show the TESS light curves and amplitude spectra of four example stars in Fig.~\ref{fig:lightcurves} to demonstrate the power of including TESS light curves in our analysis. First, is the SPB star HD\,33599 (TIC\,55295028), which has the typical signatures of multi-periodic g~modes appearing in distinct groups in its amplitude spectrum \citep{Kurtz2015,VanBeeck2021}. Its light curve also has signatures of outbursts and hence it is a candidate pulsating Be star. We also note that we have identified HD\,33599 as an SB2 system with many emission lines in its spectra, which makes it interesting to follow-up spectroscopically and understand the nature of its companion and evolutionary origin (see e.g. \citealt{Shenar2020,Bodensteiner2020}). Second, is the SPB / $\beta$\,Cep hybrid star HD\,49193 (TIC\,167523976), which has $v\sin\,i \simeq 40$\,km\,s$^{-1}$ and $T_{\mathrm{eff}} \simeq 21\,000$\,K. It is the hottest star in our sample and is located close to the cool edge of the $\beta\,$Cep instability region. 
Third is the EB system AN~Dor (TIC\,220430912), which was recently re-visited by \citet{Southworth2022} to derive masses and radii from archival RVs and new TESS light curves. Fourth, is HD\,40435 (TIC\,31313111) which has the characteristic double-wave light curve and harmonic-series in its amplitude spectrum typical of rotation modulation (see e.g. \citealt{Bowman2018}). We identified LPV in this star's spectra that is likely associated with a ApBp nature.

As a vast improvement over previous studies that utilised short TESS light curves, our re-classification has the major advantage of improved frequency resolution from long-term TESS light curves. Similar to previous work, we find a high variability fraction (> 90\%) in the TESS light curves of our sample. As mentioned previously, we do not extract pulsation, binary or rotation frequencies for follow-up modelling, as such an endeavour requires custom light curves and is the subject of future work. Our work has allowed us to rank the most promising stars with reliable spectroscopy and clear signatures of binarity and/or pulsations to follow-up with detailed seismic modelling.

\section{Discussion} \label{sect:discussion}

\subsection{Newly discovered HgMn stars} \label{sect:HgMn}
\begin{figure}
    \centering
    \includegraphics[width=\columnwidth]{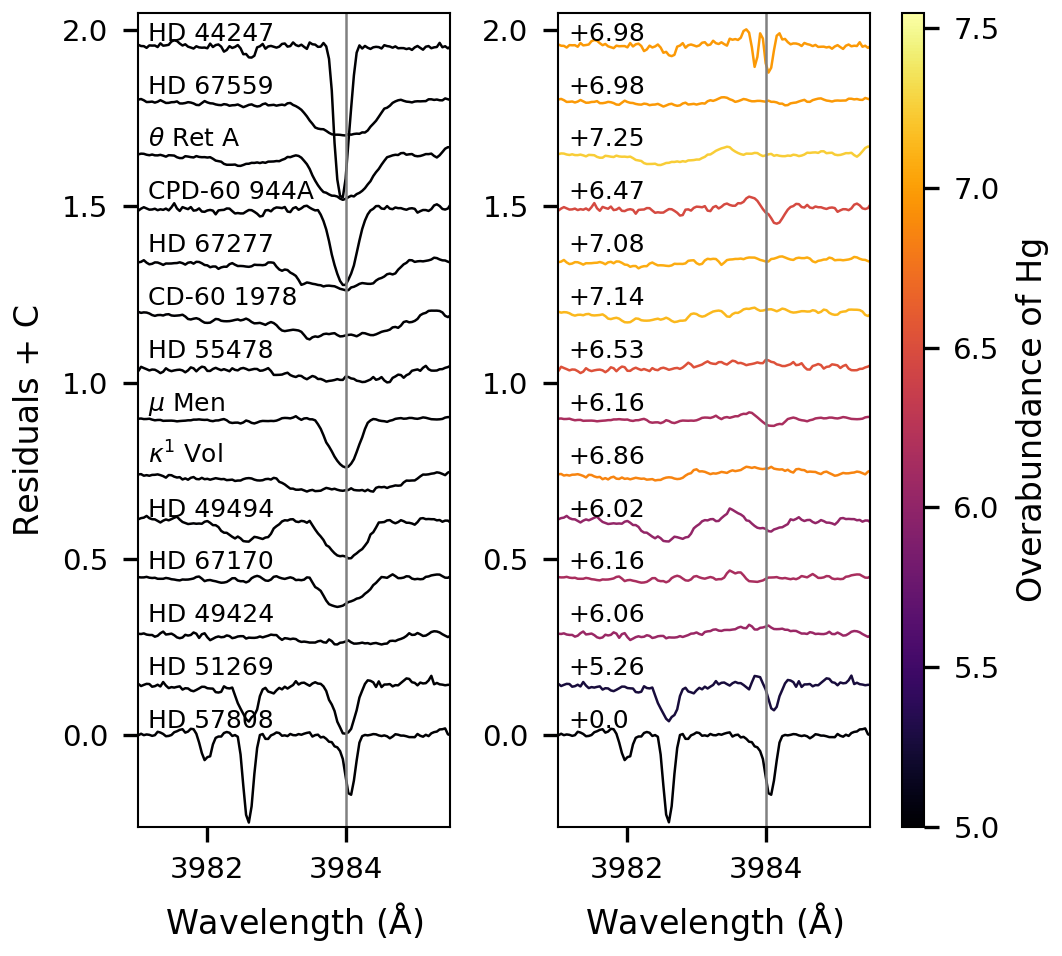}
    \caption{Overabundance of Hg in HgMn stars. \textit{Left panel} shows residuals around the \ion{Hg}{II$_{3984}$} line between the observed spectrum and best-fitting GSSP spectrum with Hg abundance according to the metallicity of the star. \textit{Right panel} shows the residuals when the Hg abundance is optimised for the star. The residuals in the right panel are colour-coded according to the Hg overabundance with respect to the abundance in the left panel. The value is also given for each star. The residuals of the different stars are offset with a constant value to plot all stars on the same figure.}
    \label{fig:Hg}
\end{figure}

\begin{figure}
    \centering
    \includegraphics[width=\columnwidth]{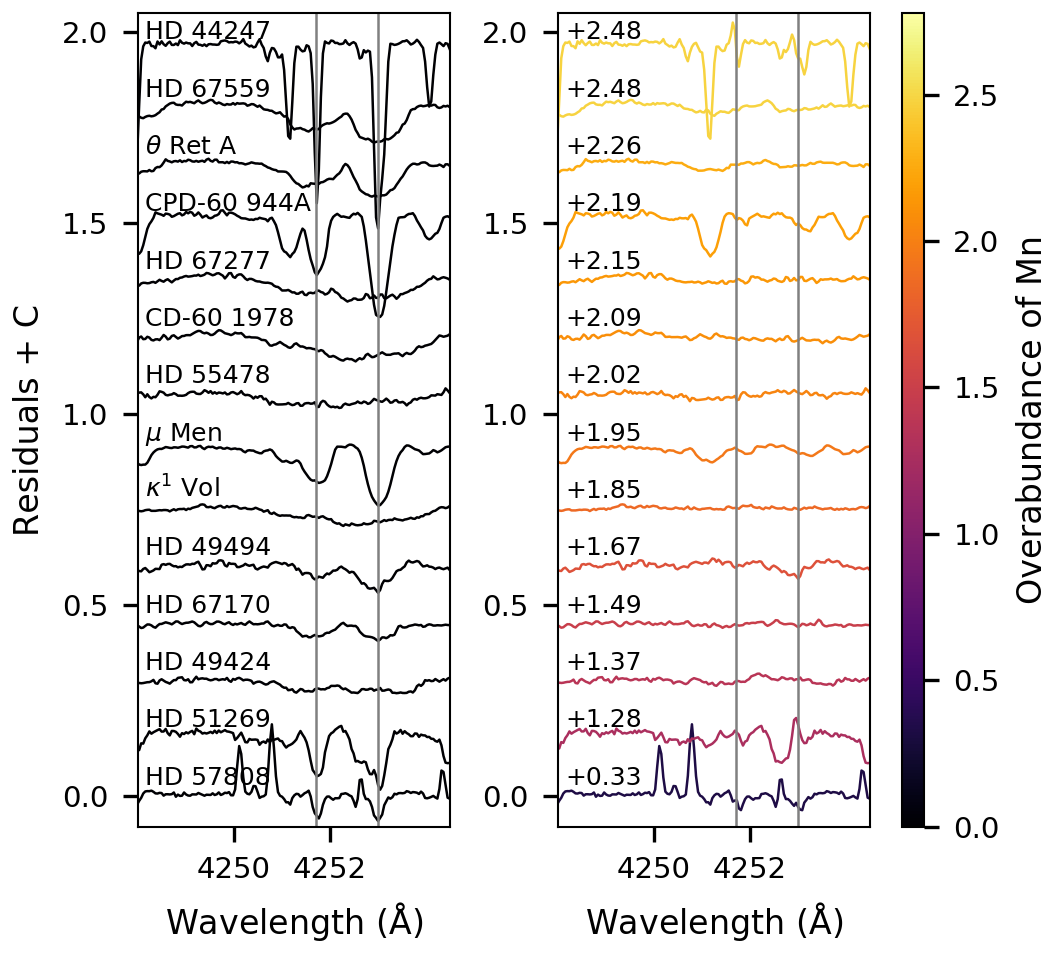}
    \caption{Overabundance of Mn in HgMn stars. Similar as in Fig.~\ref{fig:Hg} but for the \ion{Mn}{II$_{4252}$} and \ion{Mn}{II$_{4253}$} lines.}
    \label{fig:Mn}
\end{figure}

Thirteen of the main sequence stars in our sample analysed with the {\sc zeta-Payne} were found to be new HgMn stars. We optimised the Hg and Mn abundances of the 13 stars to confirm their HgMn classification. The wavelength range has been extended to 3930\,\AA\ in the blue part of the spectrum to include the \ion{Hg}{II$_{3984}$} spectral line. The differences between the observed and the model spectra around the Hg and Mn lines disappear when the abundances of these elements are optimised in the models (left versus right panels of Figs.~\ref{fig:Hg} and \ref{fig:Mn}). For some stars, such as HD\,57808, there are still some features present in the residuals after optimising for Hg and Mn. This is probably caused by overabundances of other elements such as Y, but a detailed abundance analysis is beyond the scope of this paper. 

From the TESS light curves we find that all 13 HgMn stars exhibit rotational modulation and none of them are pulsators. For a few targets also line profile variability is found in the spectra (this is indicated in column 4 of Table~A.2). 
Interestingly, only one of them is found in a SB1 binary: HD\,44247. This would result in a binarity rate of only 8\% (without correcting for observational biases) as opposed to what is found in other studies. Our observational biases include but are not limited to: an intentional bias against binary stars at the very early stage of the sample selection, exclusion of all stars that were nevertheless identified as SB2 binary systems in our sample prior to its detailed spectroscopic analysis based on the obtained FEROS spectra, and our ability to detect (especially long-period) binaries is limited by the number of acquired epochs (two in most cases), the separation of some two months between them, and the S/N in the obtained individual epoch spectra. This last bias is represented by the binary detection probabilities quoted in Sect.~\ref{sect:classification}. Therefore, the above-quoted observed binary rate among HgMn stars in our sample is unreliable.

None of these 13 stars have been classified as HgMn stars before, but $\mu\,$Men was found to have an overabundance of Si \citep{Houk1975} and CPD-60\,944A has been classified as an $\alpha^2$ CVn star with Si overabundance by \citet{Bernhard2015}.

\subsection{(Spectroscopic) Hertzsprung-Russell diagram}
\begin{figure*}
    \centering
    \includegraphics[width=\textwidth]{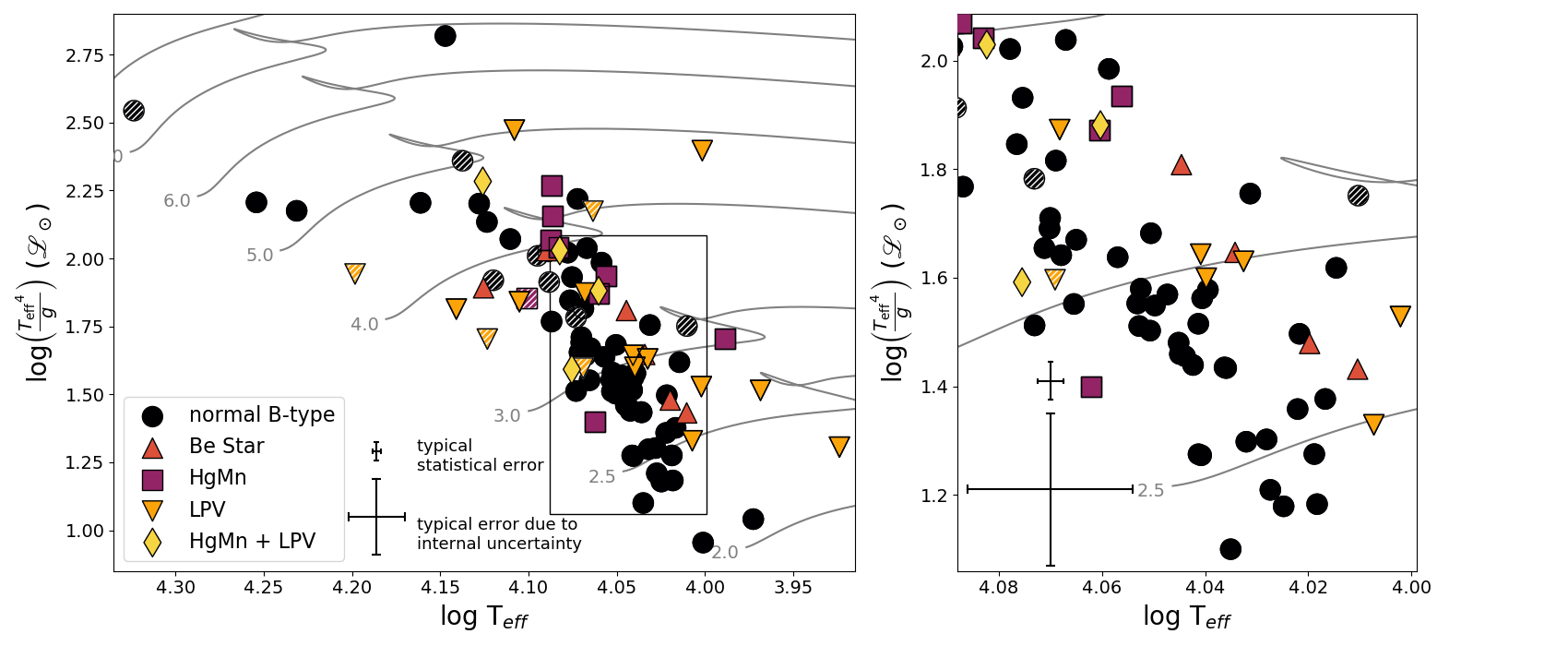}
    \includegraphics[width=\textwidth]{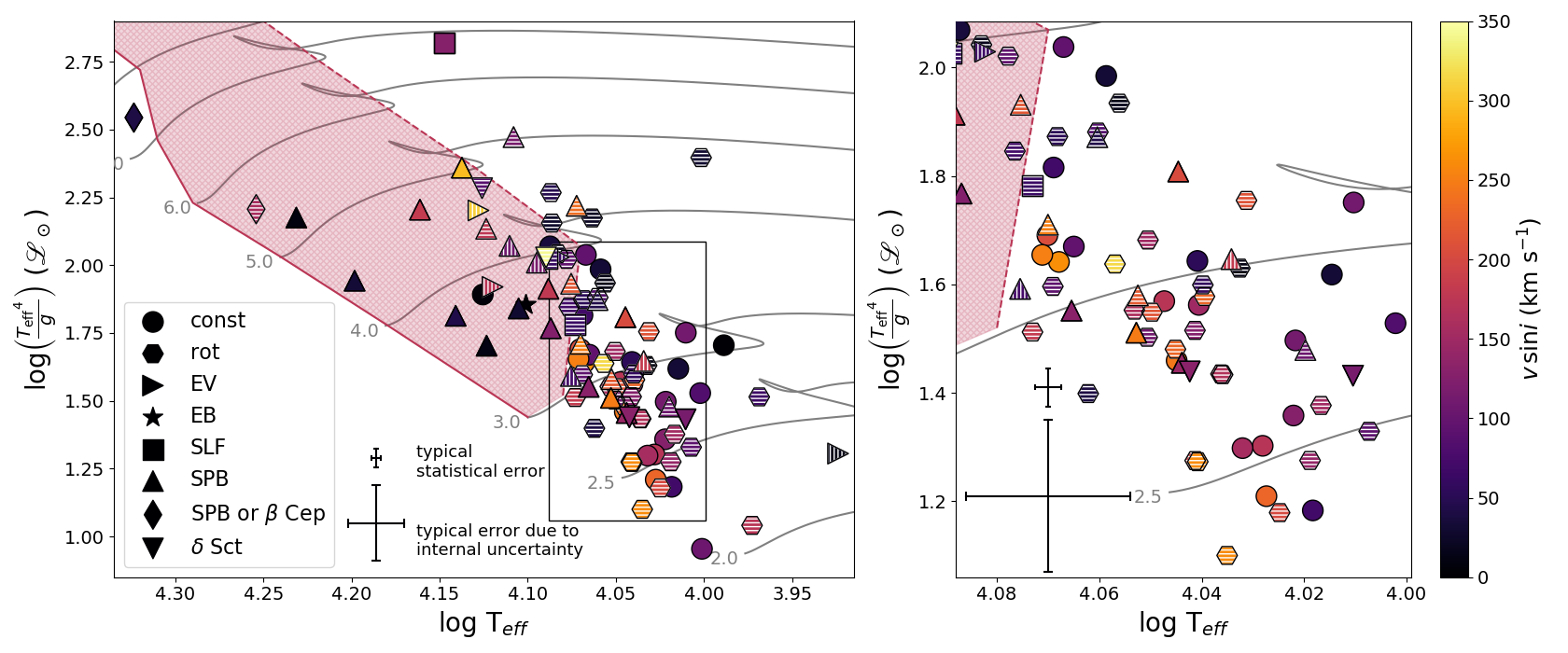}
    \caption{sHRD containing all stars from Table~A.2. \textit{Top:} the stars are colour-coded according to their spectroscopic information. The hatched symbols are SB1 binaries.
    \textit{Bottom:} same as top panel, but now the photometric variability is indicated by different symbols. They are colour-coded according to their $v\sin\,i$ values. 
    Stars with dominant rotational variability are hatched with white horizontal lines and symbols hatched with white vertical lines are stars that can either have variability according to the symbol or rotational variability. 
    The right panels are a zoom in of the region below the SPB instability strip.
    Typical errors are plotted, both the statistical error returned by the {\sc zeta-Payne} as the one including the internal uncertainties. Non-rotating MESA evolutionary tracks for different initial stellar masses (in grey) and the g-mode instability strip (red shaded area) from \citet{Burssens2020} are shown as well.}
    \label{fig:sHRD}
\end{figure*}

The stars from Table~A.2 are plotted in a spectroscopic Hertzsprung-Russell diagram (sHRD) in Fig.~\ref{fig:sHRD} in which the ordinate axis is defined as $\log\,(T_{\mathrm{eff}}^4/g)$. In the top panel, the stars are colour-coded following their spectroscopic classification, while on the bottom the symbols represent different photometric variability classes. Typical errors are shown, both for the uncertainties obtained from the {\sc zeta-Payne} and when the internal uncertainties are taken into account.
We used the non-rotating MESA evolutionary tracks for solar metallicity and an exponential core overshoot of $f_{\rm ov} = 0.02$, and the g-mode instability region from \citet{Burssens2020} to situate the stars in the sHRD. 
Most stars are intermediate-mass stars on the main sequence with masses around 2.5-4\,M$_{\odot}$. There are however two to six stars, depending on which errors are taken into account, that seem to have evolved beyond the end of the main sequence. 

\begin{figure*}
    \includegraphics[width=\textwidth]{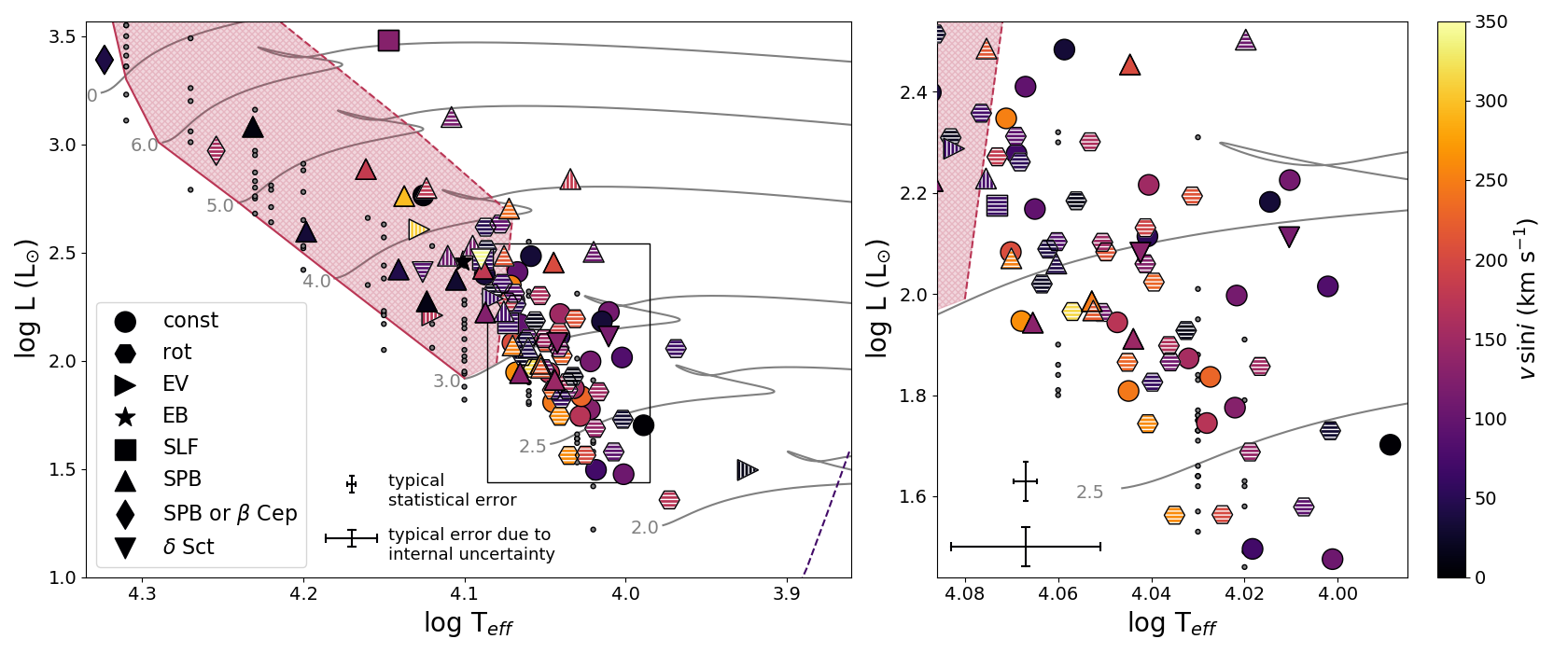}
    \caption{HRD. Same figure as the bottom panel of Fig.~\ref{fig:sHRD} but with luminosity on the y-axis. The small grey dots are B-type stars in the Scorpio-Centaurus association from \citet{Sharma2022}. Non-rotating MESA evolutionary tracks from \citet{Burssens2020} are shown in grey, and the red shaded area is the g-mode instability strip from \citet{Burssens2020}. The purple dashed line is the blue edge of the $\delta$\,Sct instability strip from \citet{Dupret2005}.}
    \label{fig:HRD}
\end{figure*}

We also included Fig.~\ref{fig:HRD} which shows the same stars in the HRD, with luminosities calculated based on their \textit{Gaia} G-band magnitudes, \textit{Gaia} eDR3 photogeometric distances from \citet{Bailer-Jones2021}, reddening values from the SFD dust map \citep{Schlafly2011} and bolometric corrections calculated with the prescriptions from \citet{Pedersen2020}. 
Apart from the g-mode instability region from \citet{Burssens2020} we also plotted the blue edge of the $\delta$\,Sct instability strip from \citet{Dupret2005}. Most of the stars classified as `SPB' or `SPB?' lie within the SPB instability strip. A large part of the sample is located outside of the SPB and $\delta$\,Sct instability regions. This includes constant stars and stars with rotational modulation, but also 10 pulsators. 
We have also added to Fig.~\ref{fig:HRD} stars from the recent work by \citet{Sharma2022}. These authors studied pulsating B-type stars in the Scorpio-Centaurus association using TESS light curves and also found pulsators below the SPB instability strip. 
Low-frequency pulsations in B-type stars cooler than the red edge of the SPB instability region have been noticed before. \citet{Salmon2014} claimed that these are fast rotating SPB stars that are seen equator-on and due to gravity darkening they appear cooler and fainter than if they would be observed pole-on. 
A similar explanation is given for high-frequency pulsators found in the literature between the red edge of the $\beta$\,Cep instability strip and the blue edge of the $\delta$\,Sct instability strip, although it is also possible that these are binaries or misclassified stars that are actually variable due to rotational modulation \citep{Balona2015,Balona2016}. 

In our sample there are eight stars classified as SPB that are situated below the SPB instability strip and that have g-mode frequencies higher than what is normally expected for these stars. 
These stars are HD\,45835, $\theta$\,Ret\,A, $\kappa$\,Men, HD\,49531, $\mu$\,Pic, HD\,47478, 29\,Dor, and HD\,37027, and none of them are spectroscopic binaries. $\mu$\,Pic, 29\,Dor and HD\,37027 are known to be Be stars. We did not include the cores of the Balmer lines in the spectral analysis for Be stars since they are in emission. Therefore the $T_{\mathrm{eff}}$ value which is derived from the Balmer lines might be subject to large uncertainties.
$\theta$\,Ret\,A is a relatively slow rotator with $v\sin\,i$ = 44 $\pm$ 13\,km\,s$^{-1}$, but is situated relatively close to the red edge of the SPB instability strip. The other seven stars are rapid rotators and might thus be fast rotating SPB stars seen equator-on as explained by \citet{Salmon2014}. Their high g-mode frequencies can be explained by fast rotation that has shifted their g-mode frequencies as observed in an inertial frame into a relatively high-frequency regime. Such an astrophysical interpretation has been proposed previously for bright SPB and Be stars observed from ground-based data \citep{Aerts1999,DeCatAerts2002,AertsKolenberg2005,Saesen2010,Saesen2013,Mowlavi2013,Mozdzierski2014,Mowlavi2016,Mozdzierski2019}, MOST space photometry \citep{Walker2005,Aerts2006,Saio2007}, CoRoT space photometry \citep{Neiner2009,Diago2009}, and {\it Kepler}/K2 space photometry \citep{Papics2017, White2017,Pedersen2021,Szewczuk2021}. 
In fast rotating stars major frequency shifts are indeed expected for prograde and retrograde sectoral modes (that is, modes with equal angular degree l and azimuthal wavenumber |m|) in the gravito-inertial regime. This is the regime where g-mode oscillations are restored by both buoyancy and the Coriolis acceleration together \citep{Townsend2005,Salmon2014,Saio2017,Aerts2019}. Our spectroscopic results and interpretation are fully in agreement with the findings by \citet{GaiaDR3} based on the \textit{Gaia} DR3 time-series photometry.

Because of the rapid rotation of these SPB stars shown by large $v\sin\,i$ values inferred from our FEROS spectroscopy, their apparently too low $T_{\mathrm{eff}}$ relative to the SPB instability region are likely to be due to the gravity darkening phenomenon. This phenomenon results from the variation of $T_{\mathrm{eff}}$ with local gravity on the surface of a non-spherical star \citep{vonZeipel1924} and therefore implies a lower $T_{\mathrm{eff}}$ at the equator than at the pole of a rapidly rotating star. As is recently demonstrated in \citet[][and private communication]{Fabry2022}, the variation of $T_{\mathrm{eff}}$ of a typical rapidly rotating B-type star amounts to about 5\% in the course of its main-sequence evolution. This $T_{\mathrm{eff}}$ variation from stellar pole to equator can explain the apparent temperature offset from the red edge of the SPB instability strip in the HRD for all but three stars. The stars for which gravity darkening is not sufficient to place them within the SPB instability strip are two Be stars HD\,37027 and 29\,Dor, and the star HD\,47478. 
However, the 5\% $T_{\mathrm{eff}}$ difference derived in \citet{Fabry2022} is based on MESA models which are limited to 1D approximations. A 2D study with the ESTER code \citep{Bouchaud2020} has shown that for the A7-type star Altair the temperature difference between equator and pole amounts to approximately 2000\,K. Thus 5\% is only a lower limit and gravity darkening effects can be higher when the proper 2D structure of fast rotators is taken into account instead of 1D approximations.

We also find four seemingly p-mode pulsators (identified as $\delta$\,Sct stars based on their TESS light curves exhibiting pulsation frequencies above $\sim 4$\,d$^{-1}$) in between the $\beta$\,Cep and $\delta$\,Sct instability regions (HD\,44533, HD\,55478, HD\,67420, and $\eta$\,Phe). These 4 stars are all fast rotators, and HD\,44533 and $\eta$\,Phe are Be stars. They are probably misclassified as $\delta$\,Sct stars, just as the high-frequency mode near $\sim 4$\,d$^{-1}$ of the rapidly rotating SPB pulsator HD\,43317 was initially thought to be a p~mode \citep{Papics2012} while it actually concerns a rotationally shifted retrograde quadrupole mode \citep{Buysschaert2018}.
In the case of the Be stars, their derived surface parameters are uncertain due to their fast rotation and circumstellar disk.

\subsection{Multiple linear regression}

We looked for correlations between all five spectroscopic parameters and the luminosity ($L$) via backward multiple linear regression.  
This was done by taking one of these six parameters as the dependent variable and performing a multiple linear regression with the other five parameters as independent variables. We computed the coefficient of determination (R$^2$) to determine how well the set of independent variables could predict the dependent variable, where R$^2$ = 1 would mean a perfect fit. For each of the independent variables we then checked if they were significant in this relation by doing a $t$-test. If the $p$-value from this $t$-test was above 0.05 the variable was assumed insignificant. The multiple linear regression was then repeated in the same way but without the least significant variable until only significant variables remained. The relations we found from this analysis are given in Table~\ref{tab:regression}.

In the first three models $\log L$, $T_{\mathrm{eff}}$, $\log\,g$ and [M/H] are related to each other. These are well-known relations expected from standard stellar structure and evolution theory. The fourth relation links $\xi$ to $\log L$, $T_{\mathrm{eff}}$, $v\sin\,i$ and [M/H]. It suggests a decrease of $\xi$ with increasing luminosity within our sample. This is in contradiction with the expectation of turbulent velocity fields getting larger as the $\log\,g$ (luminosity) value of star decreases (increases) \citep[e.g.][]{Gray2005}. This latter interpretation is supported by observational studies of single stars \citep[e.g.][]{Hunter2008} and binary systems \citep[e.g.][]{Tkachenko2014}, as well as by theoretical studies of sub-surface convective zones in intermediate- and high-mass stars \citep[e.g.][]{Cantiello2009}. We currently do not have a good physical interpretation of the (weak) opposite trend we found but we speculate it may have a methodological explanation in that $\xi$ in B-type stars cannot be reliably inferred from the global fitting procedure of their entire optical spectra. Instead, one might need to focus on a detailed analysis of carefully selected spectral lines that are sensitive to $\xi$, such as the \ion{Si}{iii} triplet at 4552~\AA, 4567~\AA, and 4574~\AA\, and/or \ion{He}{i} lines in the optical part of the spectrum. This is different from cooler stars with many metal lines for which $\xi$ can be determined from the entire spectrum since $\xi$ values from all element lines, including possible wrong values from lines that are less sensitive to $\xi$, are averaged out.

We also noticed by eye a trend of decreasing $v\sin\,i$ with age in Fig.~\ref{fig:HRD}, especially for stars with a mass below 4\,M$_{\odot}$. However, by performing a linear regression analysis for $v\sin\,i$ and stellar age, which we obtained by interpolating the MESA evolutionary tracks at the positions of the sample stars, we concluded that this correlation is not statistically significant (R$^2 = 0.02$ and R$^2 = 0.04$ for M$<$\,4\,M$_{\odot}$).

\begin{table*}
    \centering
    \caption{Multiple linear regression results}
    \begin{tabular}{lrccrcc}
    \hline\hline
    & \multicolumn{3}{c}{All variables} & \multicolumn{3}{c}{Only significant variables} \\
    \cmidrule(lr){2-4}\cmidrule(lr){5-7}
    Predictors & Estimate & p-value & R$^2$ & Estimate & p-value & R$^2$ \\
    \hline
    \multicolumn{7}{c}{Dependent variable $\log\,$L} \\
    \hline
    Intercept & 0.42(0.06) & (...) & . & 0.40(0.04) & (...) & . \\
    $T_{\mathrm{eff}}$ & 1.06(0.07) & $<$ 0.001 & . & 1.05(0.07) & $<$ 0.001 & . \\
    $\log\,g$ & -0.44(0.06) & $<$ 0.001 & . & -0.44(0.06) & $<$ 0.001 & 0.751 \\
    $v\sin\,i$ & 0.07(0.05) & 0.117 & . & & & \\ 
    $\xi$ & -0.12(0.08) & 0.104 & . & & & \\ 
    $[\mathrm{M/H}]$ & -0.06(0.07) & 0.407 & 0.761 & & & \\ 
    \hline
    \multicolumn{7}{c}{Dependent variable $T_{\mathrm{eff}}$} \\
    \hline
    Intercept & -0.15(0.06) & (...) & . & -0.18(0.04) & (...) & . \\
    $\log\,$L & 0.66(0.05) & $<$ 0.001 & . & 0.66(0.05) & $<$ 0.001 & . \\
    $\log\,g$ & 0.27(0.05) & $<$ 0.001 & . & 0.28(0.05) & $<$ 0.001 & 0.691 \\ 
    $v\sin\,i$ & -0.08(0.04) & 0.039 & . & & & \\ 
    $\xi$ & 0.09(0.06) & 0.146 & . & & & \\ 
    $[\mathrm{M/H}]$ & -0.01(0.06) & 0.857 & 0.707 & & & \\ 
    \hline
    \multicolumn{7}{c}{Dependent variable $\log\,g$} \\
    \hline
    Intercept & 0.77(0.07) & (...) & . & 0.85(0.05) & (...) & . \\
    $\log\,$L & -0.87(0.12) & $<$ 0.001 & . & -0.89(0.11) & $<$ 0.001 & . \\
    $T_{\mathrm{eff}}$ & 0.87(0.16) & $<$ 0.001 & . & 0.87(0.16) & $<$ 0.001 & . \\    
    $[\mathrm{M/H}]$ & -0.13(0.10) & 0.207 & . & -0.20(0.08) & 0.015 & 0.445 \\  
    $v\sin\,i$ & 0.08(0.07) & 0.254 & . & & & \\ 
    $\xi$ & 0.06(0.11) & 0.549 & 0.461 & & & \\ 
    \hline
    \multicolumn{7}{c}{Dependent variable $\xi$} \\
    \hline
    Intercept & 0.24(0.10) & (...) & . & 0.29(0.06) & (...) & . \\
    $\log\,$L & -0.23(0.14) & 0.104 & . & -0.29(0.11) & 0.01 & . \\
    $T_{\mathrm{eff}}$ & 0.26(0.18) & 0.146 & . & 0.32(0.16) & 0.043 & . \\    
    $v\sin\,i$ & 0.25(0.06) & $<$ 0.001 & . & 0.26(0.06) & $<$ 0.001 & . \\  
    $[\mathrm{M/H}]$ & -0.41(0.09) & $<$ 0.001 & . & -0.41(0.09) & $<$ 0.001 & 0.463 \\
    $\log\,g$ & 0.06(0.10) & 0.549 & 0.465 & & & \\ 
    \hline
    \end{tabular}
    \label{tab:regression}
\end{table*}

\section{Conclusions} \label{sect:conclusion}
We have analysed high-resolution FEROS spectra of a sample of 166 variable stars identified using TESS space photometry. Our follow-up spectroscopic data has been reduced with the CERES pipeline, to which we added four new capabilities to obtain smooth reduced normalised spectra without artificial features and cosmic hits. 

We have determined surface parameters with the {\sc zeta-Payne} framework for all Galactic single stars and SB1 systems in the sample, excluding SB2 systems and supergiants in the LMC galaxy. Apart from the uncertainties deduced with the {\sc zeta-Payne}, the internal uncertainties inherent to the neural network and fitting approach were also taken into account.
It was shown that, within the internal uncertainties, the parameters are not affected by the wavelength region that is analysed. If the neural network used in this paper is applied to longer wavelength regions, the number of coefficients for the Chebyshev polynomial representing the response function should be optimised for the respective wavelength range. Overall, in the considered wavelength interval from 4000~\AA\ to 5800~\AA\ and with 25 Chebyshev coefficients representing the residual response function of the instrument, the estimated internal uncertainties amount to some 3\% and 5\% in $T_{\mathrm{eff}}$ and $v\sin\,i$, respectively, and $\sim\,$0.1\,dex in [M/H] and $\log\,g$. These values give an upper limit on the actual parameter uncertainties of the sample stars and reach the level required for high-precision asteroseismic modelling. For this, the parameter uncertainties are ideally as small as possible but should definitely be below $\sim\,$3\% in $T_{\mathrm{eff}}$, 5-10\% in $v\sin\,i$ and $\sim\,$0.1\,dex in $\log\,g$ and for individual abundances \citep{Moravveji2016,Michielsen2021,Mombarg2022}. To obtain better constraints on the spectroscopic parameters, information from different data sources should be combined and analysed together. For example \textit{Gaia} parallaxes and magnitudes can help constrain the otherwise degenerate parameters $T_{\mathrm{eff}}$ and $\log\,g$. This is however beyond the scope of this paper and we leave this endeavour for future work.

We have found that 13 of the stars in the sample are chemically peculiar HgMn stars. The best-fitting spectra returned by the {\sc zeta-Payne} framework do not fit well the observed spectra of these stars, specifically in spectral windows that contain lines of Hg, Mn, Y, and some other chemical elements. Our ad-hoc abundance analysis of several chemical elements in the spectra of these HgMn stars with the GSSP software package revealed strong overabundances, confirming the peculiar nature of the stars.

The spectroscopic values and luminosities computed from \textit{Gaia} eDR3 data were used to place the stars in a spectroscopic and classical HRD. In these diagrams, eight stars classified as SPB with relatively high g-mode frequencies are located below the (non-rotating) SPB instability region.
In all cases the location and variability of these stars can plausibly be explained by their fast rotation, which is expected to significantly affect the instability regions of early-type stars (see \citealt{Szewczuk2017}) and shift their observed g-mode frequencies into the high-frequency regime. Therefore, we conclude to have found spectroscopic evidence of a group of fast rotating g-mode B-type pulsators monitored by TESS, covering the area between the SPB instability strip and the blue edge of the $\delta\,$Sct strip. For all but three stars, the lower temperatures relative to the SPB instability strip obtained for these stars could be related to the gravity darkening effect.

The multiple linear regression analysis involving five spectroscopic parameters and stellar luminosity revealed well-known relations between $\log L$, $T_{\mathrm{eff}}$, $\log\,g$ and [M/H] of the star as predicted by stellar evolution theory. A somewhat unexpected result is the weak negative relation found between $\log L$ and $\xi$ which we attribute to our methodology of inferring $\xi$ from the entire optical spectrum instead of selecting specific lines sensitive to $\xi$. 

Our spectrum fitting approach with the current neural network trained on LTE GSSP models is not suitable for the analysis of the spectra of supergiants. This will only be possible when NLTE models with possibility to include physics of stellar winds are used to train the neural network. Similarly, it is not yet possible to analyse SB2 systems with this version of the {\sc zeta-Payne} since this requires the analysis of either their composite or disentangled spectra. Analysis of the composite spectra of SB2 binaries is challenging and typically associated with large uncertainties in the inferred atmospheric parameters of both binary components \citep[see, e.g.][]{Tkachenko2015}. On the other hand, analysis of the disentangled spectra is as precise as spectrum analysis of spectra of single stars, provided a uniform orbital phase coverage is achieved for the SB2 system in consideration \citep[see, e.g.][]{Hadrava1995,Ilijic2004}. Therefore, detailed analysis of the spectra of evolved supergiants and of SB2 binary systems requires a dedicated effort and is beyond the scope of this paper.

\begin{acknowledgements}
The research leading to these results has (partially) received funding from the KU~Leuven Research Council (grant C16/18/005: PARADISE), from the Research Foundation Flanders (FWO) under grant agreement G0H5416N (ERC Runner Up Project), as well as from the BELgian federal Science Policy Office (BELSPO) through PRODEX grant PLATO. SG, LIJ, TVR, and DMB gratefully acknowledge support from FWO by means of PhD Aspirant mandates, and Junior and Senior Postdoctoral Fellowships, under contracts No.\,11E5620N, No.\,1124321N, No.\,12ZB620N, and No.\,1286521N, respectively.
LM thanks the European Space Agency (ESA) and BELSPO for their support in the framework of the PRODEX MAESTRO Programme.

The computational resources and services used in this work were provided by the VSC (Flemish Supercomputer Centre), funded by FWO and the Flemish Government.

This research has made use of the SIMBAD database, operated at CDS, Strasbourg, France; the SAO/NASA Astrophysics Data System; and the VizieR catalog access tool, CDS, Strasbourg, France.

The TESS data presented in this paper were obtained from the Mikulski Archive for Space Telescopes (MAST) at the Space Telescope Science Institute (STScI), which is operated by the Association of Universities for Research in Astronomy, Inc., under NASA contract NAS5-26555. Support to MAST for these data is provided by the NASA Office of Space Science via grant NAG5-7584 and by other grants and contracts. Funding for the TESS mission is provided by the NASA Explorer Program.

Some of the observations used in this work were obtained with the FEROS spectrograph attached to the 2.2-m MPG/ESO telescope at the La Silla observatory under program 0104.A-9001(A).

This work has made use of data from the ESA mission {\it Gaia} (\url{https://www.cosmos.esa.int/gaia}), processed by the {\it Gaia} Data Processing and Analysis Consortium (DPAC, \url{https://www.cosmos.esa.int/web/gaia/dpac/consortium}). Funding for the DPAC has been provided by national institutions, in particular the institutions participating in the {\it Gaia} Multilateral Agreement.

This work made use of the python packages: NumPy \citep{Numpy}, Astropy \citep{Astropy2013,Astropy2018}, SciPy \citep{Scipy}, Matplotlib \citep{Matplotlib}, pandas \citep{Pandas}, PyTorch \citep{PyTorch} and dustmaps \citep{Green2018}.

\end{acknowledgements}

\bibliographystyle{aa}
\bibliography{Gebruers2022.bib}

\end{document}